\documentclass[margin,usenatbib]{mn2e}
\usepackage{amssymb,amsmath,graphicx}
                 
\newcommand{\Jvec}{\textit{\textbf{J}} }
\newcommand{\Ivec}{\textit{\textbf{I}} }
\newcommand{\wvec}{\textit{\textbf{w}} }

\newcommand{\xvec}{\textit{\textbf{x}} }
\newcommand{\pvec}{\textit{\textbf{p}} }

\newcommand{\uvec}{\textit{\textbf{u}} }
\newcommand{\vvec}{\textit{\textbf{v}} }

\newcommand{\fvec}{\textit{\textbf{f}} }
\newcommand{\Fvec}{\textit{\textbf{F}} }
\newcommand{\gvec}{\textit{\textbf{g}} }
\newcommand{\kvec}{\textit{\textbf{k}} }

\newcommand{\Gmat}{\textsf{\textbf{G}} }
\newcommand{\Fmat}{\textsf{\textbf{F}} }
\newcommand{\Umat}{\textsf{\textbf{U}} }
\newcommand{\Smat}{\textsf{\textbf{S}} }
\newcommand{\Amat}{\textsf{\textbf{A}} }
\newcommand{\Bmat}{\textsf{\textbf{B}} }

\newcommand{\Imat}{\textsf{\textbf{I}} }
\newcommand{\Pmat}{\textsf{\textbf{P}} }
\newcommand{\Tmat}{\textsf{\textbf{T}} }

\newcommand{\Vmat}{\textsf{\textbf{V}} }

\newcommand{\bvec}{\textit{\textbf{b}} }
\newcommand{\cvec}{\textit{\textbf{c}} }
\newcommand{\dvec}{\textit{\textbf{d}} }

\newcommand{\dif}{{\rm d}}
\newcommand{\bnabla}{\mathbf{\nabla}}

\begin{document}

\title[Generalized Schwarzschild's method]{Generalized Schwarzschild's method}
\author[M.A. Jalali and S. Tremaine]
  {Mir Abbas~Jalali$^{1,2}$\thanks{mjalali@sharif.edu (MAJ)} and
   Scott Tremaine$^{2}$\thanks{tremaine@ias.edu (ST)} 
   \vspace{0.1cm} \\
   $^1$ Sharif University of Technology, Postal Code: 14588-89694, 
   Azadi Avenue, Tehran, Iran \\
   $^2$ School of Natural Sciences, Institute for Advanced Study, 
   Einstein Drive, Princeton, NJ 08540, U.S.A.}
   
\maketitle

\begin{abstract}
  We describe a new finite element method (FEM) to construct continuous
  equilibrium distribution functions of stellar systems. The method is a
  generalization of Schwarzschild's orbit superposition method from the space
  of discrete functions to continuous ones. In contrast to Schwarzschild's
  method, FEM produces a continuous distribution function (DF) and satisfies
  the intra element continuity and Jeans equations. The method employs two
  finite-element meshes, one in configuration space and one in action space.
  The DF is represented by its values at the nodes of the action-space mesh
  and by interpolating functions inside the elements. The Galerkin projection 
  of all equations that involve the DF leads to a linear system of equations, 
  which can be solved for the nodal values of the DF using linear or quadratic 
  programming, or other optimization methods. We illustrate the superior 
  performance of FEM by constructing ergodic and anisotropic equilibrium 
  DFs for spherical stellar systems (Hernquist models). We also show that 
  explicitly constraining the DF by the Jeans equations leads to smoother 
  and/or more accurate solutions with both Schwarzschild's method and FEM.
\end{abstract}


%\numberwithin{equation}{section}

\section{Introduction}
For over three decades, Schwarzschild's (1979) orbit superposition 
method has been one of the most important numerical tools for modelling 
the equilibrium states of spherical \citep{RT84}, axisymmetric \citep{T04} and triaxial 
galaxies (van den Bosch et al.\ 2008 and references therein). Schwarzschild's 
method constructs discrete phase-space distribution functions (DF) 
and works even if the gravitational potential supports chaotic orbits \citep[e.g.,][]{CD07}. 
Observational constraints can also be incorporated, in particular on the surface 
brightness or the line-of-sight velocity distributions. Schwarzschild's basic 
assumptions were: (i) the amount of mass contributed by an orbit to a 
cell/element in the configuration space is proportional to the fraction 
of time spent by that orbit inside the element; (ii) the matter density 
and velocity distribution inside each element is constant; (iii) the DF 
is non-zero only on a subset of phase space with measure zero (except 
perhaps in some cases where the potential admits large-scale chaos). 
In particular if all orbits are regular the DF is discrete, i.e., 
non-zero only at a finite set of positions in action space.
 
Despite its central role in modelling galaxies, Schwarz\-schild's 
method has several shortcomings. (i) There is no mathematical proof 
that increasing the number of elements and orbits in this scheme will 
guarantee the convergence of the coarse-grained DF to a smooth
function. (ii) In practice, Schwarzschild models often converge rather
slowly, in part because of the inverse square-root singularity in the
density contributed by an orbit near a turning point. 
(iii) Working with discrete DFs is not ideal if we require their 
derivatives for linear stability analysis, or use them to set 
initial conditions for $N$-body simulations. 

The majority of these limitations can be removed if one extends 
the method from the space of discrete functions to a more general 
continuous class. This is a problem in galactic dynamics whose 
solution is overdue and we aim to solve it using a modified 
finite element method (FEM). 

The mathematical principles of FEM are rather simple. Assume a
general governing equation 
\begin{eqnarray}
{\cal A}[u(\xvec,t) ] = 0, 
\label{eq:operator-equation}
\end{eqnarray}
for the physical quantity $u(\xvec,t)$ with ${\cal A}$ being a
partial integro-differential operator, and seek the solutions 
in terms of the coordinates $\xvec$ and the time $t$. 
An approximate solution of (\ref{eq:operator-equation}) may be 
expanded as the series
\begin{eqnarray}
u(\xvec,t) = \sum_{j=1}^{j_{\rm max}} u_j(t) \psi_j(\xvec),
\label{eq:series-for-u}
\end{eqnarray}
where $\psi_j(\xvec)$ constitute a complete set of basis functions 
that satisfy any given boundary conditions. Substituting from 
(\ref{eq:series-for-u}) into (\ref{eq:operator-equation}), multiplying
both sides of the resulting equation by $\psi_{j'}(\xvec)$, and 
integrating over the $\xvec$ domain results in
\begin{eqnarray}
\int \psi_{j'}(\xvec) ~ {\cal A}[u(\xvec,t)] ~ \dif \xvec= 0,~~
j'=1,2,\ldots,j_{\rm max}.
\label{eq:weighted-residual-for-A}
\end{eqnarray}
This is called the weighted residual form, weak form, or 
Galerkin projection of the governing equation 
(\ref{eq:operator-equation}), from which one may compute 
the unknown functions $u_j(t)$. Here, the basic assumption 
is that if the equations (\ref{eq:weighted-residual-for-A}) 
are satisfied for all $j'$ then equation (\ref{eq:operator-equation}) 
is also satisfied to an adequate degree of approximation.

It is, however, a non-trivial and sometimes impossible task to find 
suitable basis functions that (i) satisfy all boundary conditions,
(ii) make a complete set, (iii) do not contribute noise when the 
solution is changing rapidly. Moreover, the integration over the 
spatial variables is expensive if the domain of each basis function 
is the entire $\xvec$-space. One can overcome these difficulties by 
dividing the $\xvec$-space into $N$ {\it finite elements}.
The union of elements, each of volume $V_n$, is the entire $\xvec$ 
domain. Each element contains $N_{\rm d}$ nodes on its boundaries or 
in its interior. The function $u$ is approximated inside the $n$th 
element as a weighted sum of smooth shape functions $\hat\psi_{nj}(\xvec)$;
these are defined only within element $n$ and are zero at all nodes
except node $j$ of bin $n$, where they are unity. The weights $u_{nj}$
are time-dependent and are
chosen to fit the unknown value of $u$ at the nodes. The determining 
equations of the weights are
\begin{eqnarray}
&{}& \int\nolimits_{V_n} \hat \psi_{nj}(\xvec) ~ {\cal A}[u(\xvec,t)] ~ 
\dif \xvec  \nonumber \\
&{}& =\int\nolimits_{V_n} \hat \psi_{nj}(\xvec) ~ 
{\cal A}\bigg[\sum_{n,j} u_{nj}(t)\hat\psi_{nj}(\xvec) \bigg] 
~ \dif \xvec= 0, \\
&{}&\qquad  n=1,2,\ldots,N,~~j=1,\ldots,N_{\rm d}. \nonumber
\label{eq:weighted-residual-for-element}
\end{eqnarray}
This procedure, which leaves us with a system of ordinary 
integro-differential equations for the weights $u_{jk}(t)$ (the nodal values of $u$) 
is called the FEM \citep{ZT05}. 

For example, when the operator ${\cal A}$ has the form 
${\cal A}u={\partial u/\partial t}+{\cal L}u$ with ${\cal L}(t)$ a 
linear operator, the weighted residual form takes a simple matricial form 
\begin{eqnarray}
\frac{\dif}{\dif t} \uvec(t) = \Amat(t) \cdot \uvec(t) + \Fvec(t),
\label{eq:linear-reduced} 
\end{eqnarray}
where the matrix $\Amat(t)$ is the projection of the operator $-{\cal L}$ 
and $\Fvec(t)$ is a forcing vector. The vector $\uvec(t)$ contains nodal 
values of $u(\xvec,t)$. Therefore, the combination of finite elements and 
Galerkin projection reduces an infinite-dimensional partial differential 
equation to a finite-dimensional system. 

A formulation of FEM for stellar systems was presented in \cite{J10}, 
where the perturbed collisionless Boltzmann equation (CBE) was reduced 
to a form like (\ref{eq:linear-reduced}) and solved over a range of 
finite ring elements in the configuration space. That analysis, 
however, cannot be directly used to construct equilibrium DFs because 
they are not unique: according to the Jeans theorem, any distribution 
function $f$ that depends on phase-space coordinates only through the 
integrals of motion $\Ivec$ is an equilibrium solution of the CBE. Consequently
the local variation of $f$ in the $\Ivec$-space is free. 
This implies non-uniqueness and $f$ can admit discrete, piece-wise 
continuous, continuous, and differentiable solutions. 

In this paper we develop a general method to build numerical DFs
that exhibit nice properties of local differentiability and global 
continuity. After defining the problem in \S\ref{sec:equilibrium-models}, 
in \S\ref{sec:FEM-configuration-phase-space} we discuss 
finite elements, interpolation functions, and their properties both 
in the configuration and action spaces. 
In \S\ref{sec:Galerkin-weighting-fundamental-equation}, 
we obtain the Galerkin projections of velocity moments, and the 
continuity and Jeans equations. Schwarzschild's method is derived 
as a special case of FEM in \S\ref{sec:transit-time} and 
finite element models of spherical systems are discussed in 
\S\ref{sec:spherical-systems}. We apply the FEM to the 
spherical Hernquist model in \S\ref{sec:examples} and \S\ref{sec:disc}
contains a discussion of our results.

\subsection{Equilibrium stellar systems}
\label{sec:equilibrium-models}

The DF of a collisionless system in dynamical equilibrium depends 
on the position $\xvec=(x_1,x_2,x_3)$ and the velocity 
$\textbf{\textit{v}}=(v_1,v_2,v_3)$ vectors only through the 
integrals of motion. 

\paragraph*{Integrable systems} The Hamilton--Jacobi equation is
separable in spherical systems, razor-thin axisymmetric discs, 
and triaxial systems in which the potential is of St\"ackel form. 
Orbits in these systems are regular, and can be represented using 
a suitable action vector $\textbf{\textit{J}}=(J_1,J_2,J_3)$ and 
its associated angle variables $\textbf{\textit{w}}=(w_1,w_2,w_3)$. 
The Hamiltonian function ${\cal H}$ depends only on the action 
vector and the evolution of the angle variables is linear in time:
\begin{eqnarray}
\wvec(t) = {\bf \Omega} t+\wvec(0),~~
{\bf \Omega}(\Jvec)=\frac{\partial {\cal H}}{\partial \Jvec}, 
\label{eq:motion-equations-regular-orbits}
\end{eqnarray}
where ${\bf \Omega}=\left (\Omega_1,\Omega_2,\Omega_3 \right )$ is the 
vector of orbital frequencies. The actions are isolating integrals of 
motion, so any DF of the form $f(\Jvec)$ defines a possible equilibrium 
stellar system. For separable systems, the actions can be computed by 
quadratures. In this study, we focus on building equilibrium models of 
stellar systems with integrable potentials. 

\paragraph*{Non-integrable systems} Generic axisymmetric or triaxial
potentials are not integrable. There are surviving invariant tori 
of regular orbits (cf.\ KAM theory) but these are separated by chaotic 
layers. If the chaotic layers are narrow (the potential is `nearly
integrable') the DF may be assumed to be zero in the chaotic phase
subspace and written as a function of the actions in the regular phase
subspace. However, the actions must then be calculated either using
canonical perturbation theory or by generating a frequency map of 
the system \citep{L90,H02,BT08}.

A simpler and more powerful approach, which can be used even if the
potential is far from integrability, is to express the integrals of
motion in terms of initial conditions of orbits \citep{Sw79}. Thus let 
$[\xvec_0(\xvec,\vvec),\vvec_0(\xvec,\vvec)]$ be the position and
velocity of the trajectory through $(\xvec,\vvec)$ on some global
surface of section $\cal D$ through which all orbits must pass 
(e.g., a symmetry plane of a triaxial potential). Then any DF of 
the form $f(\xvec_0,\vvec_0)$ defines an equilibrium stellar system.

\subsubsection{Moments of the DF}

A collisionless system with a given density function $\rho(\xvec)$ is
a possible equilibrium if one can find a DF $f(\Jvec)\ge 0$ so that
\begin{eqnarray}
\rho(\xvec) =  \int f(\Jvec) ~ \dif \vvec.
\label{eq:density-DF-integral-form}
\end{eqnarray}
We shall also sometimes use the first- and second-order velocity moments:
\begin{eqnarray}
u^i(\xvec)  \!\!\! & \equiv & \!\!\! 
\rho \left \langle v_i \right \rangle (\xvec) =  
\int  v_i ~ f(\Jvec) ~ \dif \vvec,  \\
\tau^{ij}(\xvec) \!\!\! & \equiv & \!\!\! \rho 
\left \langle v_i v_j \right \rangle (\xvec) = 
\int  v_i v_j f(\Jvec) ~ \dif \vvec.
\label{eq:higher-moments}
\end{eqnarray}
In an equilibrium system these are related by 
the steady-state continuity equation 
\begin{eqnarray}
\sum_{i=1}^{3} \frac{\partial u^i}{\partial x_i}
= 0,
\label{eq:continuity-equation} 
\end{eqnarray}
and Jeans equations
\begin{eqnarray}
\sum_{j=1}^{3} \frac{\partial \tau^{ij} }{\partial x_j}
= -\rho \frac{\partial \Phi}{\partial x_i}, ~~ i=1,2,3,
\label{eq:Jeans-equation} 
\end{eqnarray}
with $\Phi(\xvec)$ being the potential. In systems 
with spherical symmetry only the radial and tangential velocity 
dispersions matter and three Jeans equations reduce to one. 

We shall argue below that including constraints based on the
continuity and Jeans equations can significantly improve the 
accuracy of both Schwarzschild and FEM models of stellar systems.

\section{Finite elements in configuration and action space}
\label{sec:FEM-configuration-phase-space}

\begin{figure}
\centerline{\hbox{\includegraphics[width=0.45\textwidth]{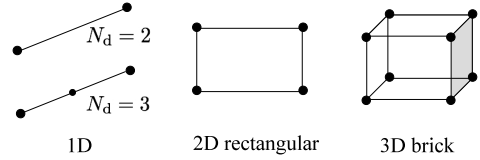} } }
\caption{Elementary finite elements. The degree of interpolating
polynomial increases by adding interior nodes, which can lie on 
the edges, sides or even inside elements.}
\label{fig1}
\end{figure}

We assume that the configuration space has been split into $N$ 
elements, each of $N_{\rm d}$ nodes, and that the density $\rho(\xvec)$ 
is known at the nodal points. Inside each element, the density function 
can be approximated by suitable interpolation (shape) functions. Denoting 
$\rho_n(\xvec)$ as the functional form of the density inside the $n$th 
element, one may write
\begin{eqnarray}
\rho(\xvec) = \sum_{n=1}^{N} H_n(\xvec) ~ \rho_n(\xvec), 
~~\rho_n(\xvec)= \sum_{k=1}^{N_{\rm d}}  
g_{k,n}(\xvec) ~ \rho_{k,n}. \label{eq:quantity-summation}
\end{eqnarray}
The density at the $k$th node of the $n$th element has been 
indexed by the pair $(k,n)$. The function $H_n(\xvec)$ is unity 
inside the $n$th element in the $\xvec$-space and zero outside. 
The interpolation functions $g_{j,n}(\xvec)$ have the following 
properties:
\begin{eqnarray}
g_{j,n} \left (\xvec_{kn} \right ) = \delta_{jk}, 
~~j,k=1,2,\ldots,N_{\rm d},
\label{eq:interpolate-Kronecker}
\end{eqnarray}
where $\delta_{jk}$ is the Kronecker delta, and $\xvec_{kn}$ 
is the position vector of the $k$th node of the $n$th element.
Figure \ref{fig1} shows some elementary one-, two- and 
three-dimensional elements. The rectangular and brick elements 
can be distorted to obtain the so-called mapped elements \citep{ZT05}, 
which help to reconstruct complex morphologies in curvilinear 
coordinates. For instance, elements confined between confocal 
ellipsoids and hyperboloids can better describe elliptical galaxy 
models that may have potentials close to St\"ackel form. Thin 
rings and spherical shells are the most efficient elements for 
axisymmetric discs and spherical systems, respectively. 

Using the superscript T to transpose a vector/matrix, 
we define the row vector 
\begin{eqnarray}
\gvec_n(\xvec) &=& \left [
\begin{array}{cccc}
  g_{1,n}(\xvec) ~ &  
~ g_{2,n}(\xvec)    
 & \cdots &
 g_{N_{\rm d},n}(\xvec) 
\end{array}
\right ], \label{eq:interpolating-vector}
\end{eqnarray}
and the column vector
\begin{eqnarray}
\bvec_n = \left [
\begin{array}{cccc}
\rho_{1,n} ~ &  ~ 
\rho_{2,n} ~ &  
\cdots       &  
\rho_{N_{\rm d},n} 
\end{array}
\right ]^{\rm T},
\label{eq:nodal-density-vector}
\end{eqnarray}
and rewrite the components of $\rho_n(\xvec)$ in the following compact form
\begin{eqnarray}
\rho_n \!\!\! & = & \!\!\! \gvec_n \cdot \bvec_n.
\label{eq:interpolate-density-compact}
\end{eqnarray}
Here a dot denotes matrix/vector multiplication. The above procedure 
can be readily applied to higher order velocity moments. In particular, 
we obtain
\begin{eqnarray} 
u^{i}_n = \gvec_n \cdot \cvec^{i}_n, ~~
\tau^{ij}_n = \gvec_n \cdot \dvec^{ij}_n,
\label{eq:interpolate-moments-compact}
\end{eqnarray}
where the column vectors $\cvec^{i}_n$ and $\dvec^{ij}_n$ contain,
respectively, the nodal values of $u^i$ and $\tau^{ij}$ inside the 
$n$th element. 

To construct a DF $f(\Jvec)$, we divide the action space to $M$ 
finite elements, each of $M_d$ nodes, and write
\begin{eqnarray}
f(\Jvec)=\sum_{m=1}^{M}  
H_m(\Jvec) ~ \textit{\textbf{f}}_m(\Jvec) \cdot \pvec_m, 
\label{eq:expand-DF-over-elements} 
\end{eqnarray}
with $\textit{\textbf{f}}_m(\Jvec)$ and $\pvec_m$ being $M_{\rm d}$ 
dimensional row and column vectors, respectively. The elements 
of the interpolating row vector $\textit{\textbf{f}}_m(\Jvec)$ are 
denoted by $f_{j,m}(\Jvec)$ and they satisfy the condition
\begin{equation}
f_{j,m}(\Jvec_{km})=\delta_{jk}, ~~ j,k=1,2,\ldots,M_{\rm d},
\end{equation}
for the action vector $\Jvec_{km}$ associated with the $k$th 
node of the $m$th element in action space. The union of the domains of $f_{j,m}$ 
covers the action space. 

In this study we use interpolation functions of $C_0$ class both 
in the $\xvec$ and $\Jvec$ spaces. The use of $C_0$ functions implies that all physical 
quantities are smooth (continuous and differentiable) inside each 
element and along its boundary lines. In the direction perpendicular 
to the boundary lines and at the nodes, the DF, density and higher 
order velocity moments will only be continuous.
For example, consider the simplest one-dimensional set of $C_0$ 
elements: element $n$ has boundaries at $x_n$ and $x_{n+1}>x_n$ 
and has two nodes, with node 1 at the smaller boundary $x_n$ and 
node 2 at the larger. The continuity of $\rho(x)$ at $x=x_{n+1}$ 
implies $\rho_n(x_{n+1})=\rho_{n+1}(x_{n+1})$. 
Using (\ref{eq:interpolate-Kronecker}), this condition reduces to
\begin{eqnarray}
b_{2,n} = b_{1,(n+1)}.
\end{eqnarray}
The first derivative of $\rho_n(x)$ with respect to $x$ exists inside 
element $n$ and is given by 
\begin{eqnarray}
\frac {\partial \rho_n}{\partial x}=b_{1,n}\frac {\partial g_{1,n}}{\partial x}
+b_{2,n}\frac {\partial g_{2,n}}{\partial x},
\end{eqnarray}
but the differentiability condition at the nodes of elements is not 
necessarily satisfied, i.e.,
\begin{eqnarray}
\left [ \frac {\partial \rho_n}{\partial x} \right ]_{x=x_{n+1}}
\not = \left [ \frac {\partial \rho_{n+1}}{\partial x} \right ]_{x=x_{n+1}}.
\end{eqnarray}
One can resolve this problem by applying $C_1$ finite elements.  
The application of $C_1$ elements requires larger vectors of nodal 
quantities (which should now include partial derivatives), and thus 
larger element matrices. In this paper we restrict ourselves to $C_0$
elements; however, we note that $C_1$ elements provide smoother solutions (at the cost of
larger matrices and greater analytic complexity), and are likely to be
useful when the partial derivatives of $f(\Jvec)$ are also present
(e.g., in linear stability analyses). 

Any vectorial function of the form $v_i^{l_1} v_j^{l_2} H_n(\xvec)\gvec_n(\xvec)$ 
can be expressed in terms of angle-action variables \citep{J10}: 
\begin{eqnarray}
v_i^{l_1} v_j^{l_2} H_n(\xvec) \gvec_n 
(\xvec) = \sum_{\kvec} 
\tilde\gvec_{\kvec }(i,j,l_1,l_2,n,\Jvec)
e^{{\rm i} \kvec \cdot \wvec },
\label{eq:Fourier-gn}
\end{eqnarray}
with $\tilde\gvec^{*}_{\kvec } = \tilde\gvec_{-\kvec }$. Here the 
asterisk stands for complex conjugation, $\textit{\textbf{k}}$ is 
a 3-vector of integers and ${\rm i}=\sqrt{-1}$. To simplify the notation, 
we will denote $\tilde\gvec_{\kvec}(i,j,l_1,l_2,n,\Jvec)$ by 
$\tilde\gvec_{\kvec}(n,\Jvec)$ if $l_1=l_2=0$, by 
$\tilde\gvec_{\kvec}(i,n,\Jvec)$ if $l_2=0$ and $l_1=1$,
and by $\tilde\gvec_{\kvec}(i,j,n,\Jvec)$ if $l_1=l_2=1$. 
These special cases correspond to the zeroth-, first-, and 
second-order velocity moments. The row vector $\tilde\gvec_{\kvec}$ 
has the same dimension as $\gvec_n$ and it is calculated from 
\begin{eqnarray}
\tilde\gvec_{\kvec } = 
\frac{1}{(2\pi)^3} \int v_i^{l_1} v_j^{l_2} ~ 
H_n(\xvec) \gvec_n(\xvec) ~
e^{-{\rm i} \kvec \cdot \wvec } ~ 
\dif \wvec.
\label{eq:Fourier-coeff}
\end{eqnarray}
When a test particle with the action vector $\Jvec$ is inside the 
$n$th element in the configuration space, the function $H_n(\xvec)$ 
is unity and that particle contributes to $\tilde\gvec_{\kvec}$. 
In other situations, the integrand of (\ref{eq:Fourier-coeff}) 
will vanish. 

To compute $\tilde\gvec_{\kvec }$, we simply integrate the equations of 
motion corresponding to the action $\Jvec$ or the initial conditions
$(\xvec_0,\vvec_0)$ until the particle enters the $n$th element at 
time $t_{1,n}$ and exits at $t_{2,n}$. We then calculate the values of 
the angles at the entry and exit times, $\wvec_{1,n}$ and 
$\wvec_{2,n}=\wvec_{1,n}+{\bf \Omega}(t_{2,n}-t_{1,n})$. We then carry out the integration
(\ref{eq:Fourier-coeff}) using Gaussian quadrature, typically with
$8$-$15$ points. The numerical integration of the equations of motion 
continues and $\tilde\gvec_{\kvec }$ is updated every time that the 
particle enters element $n$, until the trajectory closes on itself 
for periodic orbits or becomes dense in the $\wvec$-space. The only 
extra effort of this procedure compared to Schwarzschild's method
is to perform the integral (\ref{eq:Fourier-coeff}). 
In \S\ref{subsec:separable-models}, we show that one can avoid this numerical 
integration for separable models.

\section{Galerkin weighting of governing equations} 
\label{sec:Galerkin-weighting-fundamental-equation}

This section implements a Bubnov-Galerkin procedure \citep{ZT05} 
to satisfy the governing equations of physical quantities 
(dependent variables) over individual elements in a weighted residual 
sense. As a result, independent variables are eliminated from equations, 
leaving a system of algebraic equations between nodal values of DF, 
density and velocity moments. The formulation is done in Cartesian 
coordinates and it should be modified for non-Cartesian 
ones (see \S\ref{sec:spherical-systems} for spherical 
systems).

\subsection{Density and velocity moments}
\label{subsec:weighting-of-integral-equations}

Inside the $n$th element in the configuration space, equation 
(\ref{eq:density-DF-integral-form}) reduces to 
\begin{eqnarray}
H_n(\xvec) \rho_n(\xvec) = \int H_n(\xvec) f(\Jvec) ~
\dif  \vvec;
\label{eq:fundamental-equation-for-nth-element}
\end{eqnarray}
the presence of $H_n(\xvec)$ on the right side ensures that the
integration is carried out 
over a phase subspace whose particles visit the $n$th element and 
contribute to the density and velocity dispersion of that element. 
By substituting from (\ref{eq:interpolate-density-compact}) and 
(\ref{eq:expand-DF-over-elements}) into 
(\ref{eq:fundamental-equation-for-nth-element}) we obtain 
\begin{eqnarray}
H_n(\xvec) \gvec_n(\xvec) \cdot \bvec_n = 
\sum_{m=1}^{M} \! H_m(\Jvec) \! \int \! H_n(\xvec)
\fvec_m(\Jvec) \cdot \pvec_m ~\dif \vvec.
\label{eq:fundamental-equation-nth-element-expanded}
\end{eqnarray}
We now left-multiply this equation by $\dif \xvec ~ \gvec^{\rm T}_n(\xvec)$ 
and integrate the result over the $\xvec$-domain to get 
\begin{eqnarray}
\Gmat_n \cdot 
\bvec_n &=& \sum_{m=1}^{M}  \int \! \int \!\!
\dif \xvec ~ \dif \vvec ~ H_m(\Jvec) \nonumber \\
&&\times  \left [ H_n(\xvec) \gvec^{\rm T}_n(\xvec) \cdot 
\textit{\textbf{f}}_m(\Jvec) \right ]
\cdot \pvec_m,
\label{eq:fundamental-equation-Galerkin-weighting-1}
\end{eqnarray}
with 
\begin{eqnarray}
\Gmat_n = \int H_n(\xvec) \left [ \gvec^{\rm T}_n(\xvec)
\cdot \gvec_n(\xvec) \right ] \dif \xvec,
\label{eq:define-matrix-G}
\end{eqnarray}
being an $N_{\rm d}\times N_{\rm d}$ constant matrix. The function 
$H_n(\xvec)$ in the integrand of (\ref{eq:define-matrix-G}) restricts 
the domain of integration to the region occupied by the $n$th element. 
The integral in $\Gmat_n$ can be done analytically should one use 
interpolation functions of polynomial type. 

The transformation $(\xvec,\vvec) \rightarrow (\wvec,\Jvec)$ is 
canonical and so the infinitesimal phase space volume $\dif\xvec ~\dif\vvec$ 
can be replaced by $\dif\wvec ~\dif\Jvec$. Using (\ref{eq:Fourier-gn}) 
with $l_1=l_2=0$, equation (\ref{eq:fundamental-equation-Galerkin-weighting-1}) 
is transformed to
\begin{eqnarray}
\bvec_n \!\!\! &=& \!\!\! \sum_{m=1}^{M} \sum_{\kvec }  
\int \!\! \int 
\dif\wvec ~ \dif\Jvec ~ H_m(\Jvec) ~
e^{ {\rm i} \kvec \cdot \wvec } \nonumber \\
&{}& \qquad \times \Gmat^{-1}_n \cdot \left [ 
\tilde\gvec^{\rm T}_{\kvec}(n,\Jvec) \cdot 
\textit{\textbf{f}}_m(\Jvec) \right ]
\cdot \pvec_m.
\label{eq:Galerkin-action-angle-1}
\end{eqnarray}
It is obvious that only the term with
$\textit{\textbf{k}}=(0,0,0)\equiv \textit{\textbf{0}}$ contributes to
the integral over the $\wvec$-space and equation
(\ref{eq:Galerkin-action-angle-1}) reads
\begin{eqnarray}
\bvec_n = \sum_{m=1}^{M} \Fmat_{\rm e}(n,m) \cdot 
\pvec_m,~~n=1,2,\ldots,N,
\label{eq:Galerkin-weighted-final-1}
\end{eqnarray}
with
\begin{eqnarray}
\Fmat_{\rm e}(n,m) = (2\pi)^3 \!\! \int \dif \Jvec ~ H_m(\Jvec)~
\Gmat^{-1}_n \cdot \left [ \tilde\gvec^{\rm T}_{
\textit{\textbf{\scriptsize{0}}} }(n,\Jvec) \cdot
\textit{\textbf{f}}_m(\Jvec) \right ].
\label{eq:element-equation-Fe}
\end{eqnarray}
Repeating the above procedure for the functions $H_n(\xvec) u^i(\xvec)$ and
$H_n(\xvec)\tau^{ij}(\xvec)$ leads to
\begin{eqnarray} 
\cvec^i_n =\sum_{m=1}^{M} \Umat_{\rm e}(i,n,m) \cdot 
\pvec_m,~~ \dvec^{ij}_n =\sum_{m=1}^{M} \Smat_{\rm e}(i,j,n,m) \cdot 
\pvec_m, \label{eq:Galerkin-weighted-final-2}
\end{eqnarray}
where
\begin{eqnarray}
\Umat_{\rm e} \!\!\! & = & \!\!\! (2\pi)^3 \!\! \int \dif \Jvec ~ H_m(\Jvec) ~
 \Gmat^{-1}_n \cdot \left [ \tilde\gvec^{\rm T}_{
\textit{\textbf{\scriptsize{0}}} }(i,n,\Jvec) \cdot
\textit{\textbf{f}}_m(\Jvec) \right ],  
\label{eq:element-equation-Ue} \\
\Smat_{\rm e} \!\!\! & = & \!\!\! (2\pi)^3 \!\! \int \dif \Jvec ~ H_m(\Jvec) ~
\Gmat^{-1}_n \cdot \left [ \tilde\gvec^{\rm T}_{
\textit{\textbf{\scriptsize{0}}} }(i,j,n,\Jvec) \cdot
\textit{\textbf{f}}_m(\Jvec) \right ].
\label{eq:element-equation-Se}
\end{eqnarray}
The constant element matrices $\Fmat_{\rm e}$, $\Umat_{\rm e}$ and $\Smat_{\rm e}$ 
are of the size $N_{\rm d}\times M_{\rm d}$ and there are $N\times M$ of 
them. There are additional constraints associated with the 
{\it element equations} (\ref{eq:Galerkin-weighted-final-1}) and 
(\ref{eq:Galerkin-weighted-final-2}) at a node shared by several elements, 
since a physical quantity must have the same value in the Galerkin projections 
of all those elements. In fact, one can introduce $N_{\rm t}$-dimensional 
column vectors $\bvec$, $\cvec_i$ and $\dvec_{ij}$ that contain all nodal 
densities and first- and second-order velocity moments, and because nodes are 
shared the dimension $N_{\rm t} < N\times N_{\rm d}$. Similarly, the nodal 
DFs constitute an $M_{\rm t}$-dimensional column vector $\pvec$ where 
$M_{\rm t}<M\times M_{\rm d}$ is the total number of distinct nodes in 
the $\Jvec$-space. Equation (\ref{eq:Galerkin-weighted-final-1}) can thus
be written as 
\begin{eqnarray}
\bvec = \Fmat \cdot \pvec.
\label{eq:assembled-linear-system-a}
\end{eqnarray}
This matrix equation can be solved to yield the DF, as parametrized
by its nodal values $\pvec$. The rank of the matrix $\Fmat$ is
generally less than its dimension $N_{\rm t}$, and there are
additional constraints that the DF should be non-negative, so in
either Schwarzschild's method or the FEM the solution must be sought
by linear or quadratic programming or some other optimization method
(see \S\ref{eq:linear-programming}). Once the solution is known, the
nodal values of the streaming velocity and velocity-dispersion tensor 
are obtained from equations (\ref{eq:Galerkin-weighted-final-2}), 
which can be written as
\begin{eqnarray}
\cvec_{i}=\Umat(i) \cdot \pvec,
~~ \dvec_{ij}=\Smat(i,j) \cdot \pvec.
\label{eq:assembled-linear-system-b}
\end{eqnarray}
The $N_{\rm t}\times M_{\rm t}$ constant global matrices $\Fmat$,
$\Umat$ and $\Smat$ are generally dense. Assembling the element equations is a 
routine procedure in finite element analysis, and its logic is 
to use the continuity condition and eliminate repeated nodal 
quantities (like density and DF) from all element equations 
except one. 

\subsection{Continuity and Jeans equations}

\label{sec:jeans}

The accuracy of FEM models of equilibrium stellar systems can be
improved by adding additional constraints that ensure that the
continuity and Jeans equations (\ref{eq:continuity-equation}) and
(\ref{eq:Jeans-equation}) are satisfied. Inside the $n$th element,
these equations can be written
\begin{eqnarray}
H_n(\xvec) \sum_{j=1}^{3} \frac{\partial \gvec_n }{\partial x_j}
\cdot \cvec^{j}_n
 \!\!\! & = & \!\!\! 0, 
 \label{eq:element-continuity-1} \\
H_n(\xvec) \sum_{j=1}^{3} \frac{\partial \gvec_n }{\partial x_j}
\cdot \dvec^{ij}_n
\!\!\! & = & \!\!\!  
-H_n(\xvec) \frac{\partial \Phi}{\partial x_i}
\gvec_n \cdot \bvec_n, ~~ i=1,2,3.
\label{eq:element-Jeans-2} 
\end{eqnarray}
Defining 
\begin{eqnarray}
\Tmat^j_n &=& \int H_n(\xvec) \left [ \gvec_n^{\rm T} 
\cdot \frac{\partial \gvec_n }{\partial x_j} \right ] ~\dif \xvec,
\label{eq:coeff-Jeans-equation-element-1} \\
\Umat_n(i) &=& -\int 
H_n(\xvec) ~ \frac{\partial \Phi}{\partial x_i} ~
\left [ \gvec_n^{\rm T}\cdot \gvec_n \right ] 
~ \dif \xvec, \label{eq:coeff-Jeans-equation-element-2}
\end{eqnarray}
one obtains the Galerkin projections of (\ref{eq:element-continuity-1})
and (\ref{eq:element-Jeans-2}) as
\begin{eqnarray}
\sum_{j=1}^{3} \Tmat^j_n \cdot \cvec^{j}_n =0,~~
\sum_{j=1}^{3} \left [ \Umat_n(i) \right ]^{-1} \cdot 
\Tmat^j_n \cdot \dvec^{ij}_n =  \bvec_n.
\label{eq:Jeans-equation-element} 
\end{eqnarray}
We assemble these element equations to obtain the global forms
\begin{eqnarray}
\sum_{j=1}^{3} \Amat(j) \cdot \cvec_{j} = 0, ~~
\sum_{j=1}^{3} \Bmat(j) \cdot \dvec_{ij} = \bvec.
\label{eq:Jeans-equation-assembled} 
\end{eqnarray}
Combining (\ref{eq:assembled-linear-system-b}) and 
(\ref{eq:Jeans-equation-assembled}) leads to 
\begin{eqnarray}
\sum_{j=1}^{3} \left [ ~ \Amat(j) \cdot \Umat(j) 
~ \right ] \cdot \pvec = 0, ~~
\sum_{j=1}^{3} \left [ ~ \Bmat(j) \cdot \Smat(i,j) 
~ \right ] \cdot \pvec = \bvec.
\label{eq:Jeans-equation-assembled-vs-DFs} 
\end{eqnarray}
We make some remarks. (i) The solutions of the continuity and Jeans
equations, whether the continuous versions
(\ref{eq:continuity-equation}) and (\ref{eq:Jeans-equation}) or their
FEM counterparts (\ref{eq:Jeans-equation-assembled}) above, are
generally not unique.  A notable exception is triaxial potentials of
St\"ackel form, in which the second-order tensor $\tau^{ij}(\xvec)$ is
diagonal in ellipsoidal coordinates \citep{VHZ03}. 
(ii) When using $C_0$ finite elements, as we do here, the moments
$\rho$, $u^i$, and $\tau^{ij}$ (eqs.\
\ref{eq:density-DF-integral-form}--\ref{eq:higher-moments}) are
continuous across element boundaries but generally their derivatives
are not; however, since the right-hand sides of the continuity and
Jeans equations {\it are} continuous across boundaries, the combinations of
derivatives of $u^i$ and $\tau^{ij}$ that appear on the left-hand
sides of these equations must also be continuous.  (iii) Equations
(\ref{eq:Jeans-equation-assembled-vs-DFs}) do not say that the mass
and momentum flows into each element, through its boundaries, exactly
balance their outflows (although the balance will become more and more
accurate as the number of nodes increases).  Only finite volume
methods \citep{LeVeque} and conservative FEMs ensure the exact
conservation of physical fluxes and this paper does not discuss those techniques.

\subsection{Separable models}
\label{subsec:separable-models}

The computation of the element matrices $\Fmat_{\rm e}$, $\Umat_{\rm e}$
and $\Smat_{\rm e}$ is accelerated when the Hamilton--Jacobi equation
is separable for the potential $\Phi(\xvec)$. In such a circumstance,
the velocity vector can be expressed as $\vvec(\xvec,\Jvec)$, which  
implies 
\begin{eqnarray}
\dif \vvec = {\cal Q}(\xvec,\Jvec) ~ \dif \Jvec,~~
{\cal Q}(\xvec,\Jvec) = \frac{\partial(v_1,v_2,v_3)}{\partial (J_1,J_2,J_3)}.
\end{eqnarray}
This allows us to bypass the costly integration of orbit 
equations needed for calculating the Fourier coefficients 
$\tilde \gvec_{\textit{\textbf{\scriptsize{0}}} }$. Defining 
the matrix
\begin{eqnarray}
\Pmat(\xvec,\Jvec) = H_n(\xvec) H_m(\Jvec) ~ 
{\cal Q}(\xvec,\Jvec) ~ \Gmat^{-1}_n \cdot 
\left [ \gvec^{\rm T}_n(\xvec) \cdot \textit{\textbf{f}}_m(\Jvec) \right ],
\end{eqnarray}
the element matrices are computed from
\begin{eqnarray}
\Fmat_{\rm e} \!\!\! & = & \!\!\! \int\int  
~ \Pmat(\xvec,\Jvec) ~ \dif \xvec ~ \dif \Jvec,  
\label{eq:matrix-F} \\
\Umat_{\rm e} \!\!\! & = & \!\!\! \int\int  
~ v_i(\xvec,\Jvec) ~ \Pmat(\xvec,\Jvec) ~ \dif \xvec ~ \dif \Jvec, 
\label{eq:matrix-U} \\
\Smat_{\rm e} \!\!\! & = & \!\!\! \int\int  
~ v_i(\xvec,\Jvec) ~ v_j(\xvec,\Jvec) ~ \Pmat(\xvec,\Jvec) 
~ \dif \xvec ~ \dif \Jvec.
\label{eq:matrix-S}
\end{eqnarray}
The integrals over the $\xvec$ and $\Jvec$ subdomains can be 
evaluated using Gaussian quadratures. 

In separable models the turning points of orbits and their shapes 
in the configuration space are known. Therefore, the null integrals 
in the element matrices can be avoided, and the computational effort 
is reduced, by a priori identification of the $\Jvec$-subspaces whose 
orbits never enter a selected element in the $\xvec$-space. In fact, 
the information related to the passage of an orbit through a given 
element is carried by the function $H_m(\Jvec)H_n(\xvec)$, and before evaluating the 
integrals we can exclude all $(m,n)$ pairs that give 
$H_n(\xvec)H_m(\Jvec)=0$. For separable models in non-Cartesian 
coordinates, the velocity components in the Jacobian ${\cal Q}$ 
are replaced by generalized momenta, and the matrix $\Pmat$ is 
divided by the product of metric coefficients.

\subsection{Linear and quadratic programming}
\label{eq:linear-programming}

The size of the unknown vector $\pvec$ is not necessarily, or usually, 
equal to the total number of constraints. Even if it were, the solution 
vector would not necessarily fulfill the requirement that the DF must 
be non-negative. We therefore employ either linear programming (LP) 
or quadratic programming (QP; \citealt{G81}) and search for $p_l$, 
the components of $\pvec$, by minimizing the objective function
\begin{eqnarray}
{\cal J}=\sum_{l=1}^{M_{\rm t}} 
C_l p_l+\frac {1}{2} \sum_{l=1}^{M_{\rm t}}\sum_{l'=1}^{M_{\rm t}}W_{ll'} ~ 
p_l p_{l'} ,
\label{eq:objective}
\end{eqnarray}
with $W_{ll'}=0$ for LP. The minimization is 
subject to the inequalities $p_l\ge 0$ (for $l$=$1,2,\ldots,M_{\rm t}$) and
the equality constraints (\ref{eq:assembled-linear-system-a}).  If we also
demand satisfaction of the continuity and Jeans equation these are
supplemented by the equality constraints
(\ref{eq:Jeans-equation-assembled-vs-DFs}).  The QP routines begin from a
vector $\pvec_0$ that satisfies the imposed constraints with a tolerance of
$\epsilon_f$, then proceed to minimize ${\cal J}$. The vector $\pvec_0$ is
usually called a feasible solution and $\epsilon_f$ is the feasibility
tolerance; the latter must be greater than the computational accuracy of
the variables involved in the constraints.

The weights $W_{ll'}$ are chosen based on the desired attributes of the
model, such as bias toward radial or tangential orbits, maximization
of a quadratic entropy, or fitting to specified data. For example, if
a series of observables $o_\alpha$ are linear functions of the DF,
\begin{equation}
o_\alpha=\sum_{l=1}^{M_{\rm t}} O_{\alpha l}p_l, \quad \alpha=1,\ldots,K,
\end{equation}
and they are observed to have values $\overline o_\alpha$ with
observational errors $\sigma_\alpha$, then a suitable objective
function is specified by
\begin{equation}
C_l=-\sum_{\alpha=1}^K\overline o_\alpha O_{\alpha l}, \quad
W_{ll'}=\sum_{\alpha=1}^{K} \frac{O_{\alpha l} O_{\alpha l'}}{\sigma_\alpha^2}.
\end{equation}
The LP and QP algorithms we have used can stall at weak local minima or ``dead
points''.  Whenever this happens, we perturb the solution and restart the
algorithm. 

\section{Derivation of Schwarzschild's method}
\label{sec:transit-time}

It is now straightforward to show that Schwarzschild's orbit superposition
method is a subclass of FEM. We assume for simplicity that the potential is
integrable so the orbits are regular. The orbit library in Schwarzschild's
method is collected by sampling over the space of initial conditions. For
regular orbits there is a one to one and invertible map between
$(\xvec_0,\vvec_0)$ and $[\wvec(0),\Jvec]$, and Schwarzschild's DFs will have the
following form \citep{V84}
\begin{eqnarray}
f(\Jvec)=\frac{{\cal M}}{(2\pi)^3}\sum_{m=1}^{M} p_m \delta(\Jvec-\Jvec_m),
\label{eq:DF-vs-delta-functions}
\end{eqnarray}
where $\delta(\cdots)$ is the Dirac delta function and $p_m$ is the 
discrete DF associated with an orbit of the action vector $\Jvec_m$.
Equation (\ref{eq:DF-vs-delta-functions}) is derived from 
(\ref{eq:expand-DF-over-elements}) by shrinking the elements
in the action space to zero size. The total mass of the galaxy is 
computed from 
\begin{eqnarray}
{\cal M}=\int \!\! \int f(\Jvec) ~ \dif \Jvec ~ \dif \wvec,
\end{eqnarray}
which is combined with (\ref{eq:DF-vs-delta-functions}) 
to obtain the constraint
\begin{eqnarray}
\sum_{m=1}^{M} p_m =1.
\label{eq:constraint-DF-unity}
\end{eqnarray}

Schwarzschild assumes a uniform density $\rho_n$ inside the $n$th element 
in configuration space. This implies that there is one node per
element ($N_{\rm d}=1$) and that the vector function $\gvec_n(\xvec)$ 
reduces to a scalar constant, $g_n=1$. Equation (\ref{eq:quantity-summation}) 
then reduces to 
\begin{eqnarray}
\rho(\xvec)=\sum_{n=1}^{N} H_n(\xvec) ~ \rho_n.
\label{eq:unifrom-density}
\end{eqnarray}
The matrix $\Gmat_n$ is now a single number $V_n$, which is the volume of 
the $n$th element. The quantity ${\cal M}_n$=$\Gmat_n \cdot \bvec_n$=$V_n \rho_n$ 
will thus be the mass inside the $n$th element. We substitute 
(\ref{eq:DF-vs-delta-functions}) and (\ref{eq:unifrom-density}) into 
(\ref{eq:Galerkin-weighted-final-1}) and obtain
\begin{eqnarray}
{\cal M}_n = {\cal M} \sum_{m=1}^{M} p_m 
~ \tilde g_{\textit{\textbf{\scriptsize{0}}} }(n,\Jvec_m),
\label{eq:Galerkin-weighted-delta-DF}
\end{eqnarray}
with the zeroth-order Fourier coefficient given by
\begin{eqnarray}
\tilde g_{\textit{\textbf{\scriptsize{0}}} }(n,\Jvec_m)
=\frac{1}{(2\pi)^3} \int H_n[\xvec(\wvec,\Jvec_m)] ~ \dif \wvec.
\label{eq:g0-for-uniform-density-element}
\end{eqnarray}
According to time averages theorem \citep{BT08}, the quantity 
on the right hand side of (\ref{eq:g0-for-uniform-density-element})
is the fraction of time $t_n(\Jvec_m)$ that an orbit of action 
$\Jvec_m$ spends in the $n$th spatial element. Consequently, we obtain 
\begin{eqnarray}
{\cal M}_n = {\cal M} \sum_{m=1}^{M} t_n(\Jvec_m) ~ p_m,
\label{eq:Schwarzschild-equation}
\end{eqnarray}
which is Schwarzschild's equation. 

The approach described in \S\ref{sec:jeans} to incorporate constraints
based on the continuity and Jeans equations into FEM models does not
work for Schwarzschild's method: because the interpolating functions
$\gvec$ are constants, the matrices $\Tmat_n^j$ defined in equation
(\ref{eq:coeff-Jeans-equation-element-1}) are zero so the first of
equations (\ref{eq:Jeans-equation-element}) is trivially satisfied
and the second has no solution. Physically, the stress tensor
$\tau^{ij}_n$ is constant within an element so there is no divergence
in the momentum flux to balance the gravitational force per unit
volume $-\rho\partial\Phi/\partial x_i$.

The failure of Schwarzschild's method to satisfy the Jeans equations
within an element does not imply that the method is incorrect. Indeed,
the method {\it must} satisfy the Jeans equations on larger scales
because the assumed discrete DF (\ref{eq:DF-vs-delta-functions})
depends only on the actions and hence must satisfy the CBE, and the
Jeans equations are moments of the CBE. The correct statement is that
Schwarzschild's method satisfies the Jeans equations approximately if 
we calculate gradients of the stress tensor {\it between} adjacent
elements and match their sum to $-\rho\bnabla\Phi$ at the center of an
element. In this process, which we carry out for spherical systems 
in \S\ref{sec:spherical-systems}, one must appropriately handle 
partial derivatives because elements in the configuration space 
are not usually separated uniformly.

\section{Spherically symmetric systems}
\label{sec:spherical-systems}

In spherical systems one can use simple shell elements 
(ring elements for axisymmetric disks). Moreover, the 
distance of particles from the centre is represented as the 
Fourier series of the radial angle $w_R$ only. This remarkably 
simplifies the calculation of the vectorial function 
$\tilde \gvec_{\textit{\textbf{\scriptsize{0}}} }$ should one 
decide to compute the element matrices from 
(\ref{eq:element-equation-Fe}), (\ref{eq:element-equation-Ue})
and (\ref{eq:element-equation-Se}). Consider the spherical 
polar coordinates $(r,\theta,\phi)$ and the corresponding 
velocities $(v_r,v_{\theta},v_{\phi})$ where $r$ is the 
radial distance from the centre, $\theta$ is the co-latitude 
and $\phi$ is the azimuthal angle. The density of a spherical 
system is a function of $r$, its first-order velocity moment 
$\langle v_r \rangle$ is zero, and the following relations 
hold for second-order velocity moments:  
\begin{eqnarray}
\langle v_t^2 \rangle = 2 \langle v_{\phi}^2 \rangle = 
2 \langle v_{\theta}^2 \rangle,~~
\langle v_r v_{\phi} \rangle = 
\langle v_r v_{\theta} \rangle =0,
\end{eqnarray}
where $v_t^2=v_{\theta}^2+v_{\phi}^2=L^2/r^2$
with $L$ being the magnitude of angular momentum vector. 
We confine ourselves to models with 
$\langle v_{\phi} \rangle$=$\langle v_{\theta} \rangle$=$\langle v_{\phi} v_{\theta} \rangle$=0.
The continuity equation (mass conservation) is trivially satisfied for 
such a system. The elements of the stress tensor are
\begin{eqnarray}
\tau^{rr}=\rho \langle v_r^2 \rangle,~~
\tau^{tt}=\rho \langle v_t^2 \rangle,~~
\tau^{\theta\theta}=\rho \langle v_{\theta}^2 \rangle,~~
\tau^{\phi\phi}=\rho \langle v_{\phi}^2 \rangle,
\end{eqnarray}
and the Jeans equation in the radial direction reads
\begin{eqnarray}
\frac{\dif \tau^{rr}}{\dif r}+\frac{1}{r}\left ( 2\tau^{rr}-
\tau^{tt} \right ) = -\rho \frac{\dif \Phi}{\dif r}.
\label{eq:Jeans-radial-sphr}
\end{eqnarray}
The other two equations, in the $\theta$- and $\phi$-direction, 
do not provide further information.

We consider a mesh of $N$ concentric shell elements and define 
the $n$th element by its inner and outer radii $r_n$ and $r_{n+1}$, 
respectively. We then approximate the density and velocity moments 
as
\begin{eqnarray}
\rho(r) \!\!\! &=& \!\!\! 
\sum_{n=1}^{N} H_n(r) \gvec_n(r) \cdot \bvec_n, 
\label{eq:density-sphr-r} \\
\tau^{rr}(r) \!\!\! &=& \!\!\! 
\sum_{n=1}^{N} H_n(r) \gvec_n(r) \cdot \dvec^{rr}_n, 
\label{eq:vrr-sphr-r} \\
\tau^{tt}(r) \!\!\! &=& \!\!\! 
\sum_{n=1}^{N} H_n(r) \gvec_n(r) \cdot \dvec^{tt}_n.
\label{eq:vtt-sphr-r}
\end{eqnarray}
The continuity conditions at the boundaries of adjacent elements are
$b_{1,(n+1)}=b_{N_{\rm d},n}$, 
$d^{rr}_{1,(n+1)}=d^{rr}_{N_{\rm d},n}$ and 
$d^{tt}_{1,(n+1)}=d^{tt}_{N_{\rm d},n}$. Let us now substitute 
from equations (\ref{eq:density-sphr-r})--(\ref{eq:vtt-sphr-r}) 
into (\ref{eq:Jeans-radial-sphr}) and derive its Galerkin 
projection as
\begin{eqnarray}
\sum_{m=1}^{M} \Vmat_n^{-1} \cdot 
\left [ \Tmat^{r}_n \cdot \Smat^{rr}_{\rm e}(n,m) 
- \Tmat^{t}_n \cdot \Smat^{tt}_{\rm e}(n,m) \right ]\cdot \pvec_m =
\bvec_n, 
\label{eq:element-based-jeans}
\end{eqnarray}
for $n=1,2,\ldots,N$. The matrices $\Smat^{rr}_{\rm e}$ 
and $\Smat^{tt}_{\rm e}$ are determined from (\ref{eq:matrix-S}) 
by replacing $v_i v_j$ with $v_r^2$ and $v_t^2$, respectively, 
and we have 
\begin{eqnarray}
\Tmat^{r}_n \!\!\! &=& \!\!\! \int H_n(r) 
\left [ \gvec_n^{\rm T} \cdot \frac{\dif \gvec_n }{\dif r}
\right ] ~ r^2\dif r + 2 \Tmat^{t}_n,
\label{eq:coeff-Jeans-sphr-1} \\
\Tmat^{t}_n \!\!\! &=& \!\!\! \int H_n(r) 
\left [ \frac {1}{r} \gvec_n^{\rm T} \cdot \gvec_n \right ] 
~ r^2 \dif r,
\label{eq:coeff-Jeans-sphr-2} \\
\Vmat_n \!\!\! &=& \!\!\! -\int 
H_n(r) ~ \frac{\dif \Phi}{\dif r} 
\left [ \gvec_n^{\rm T} \cdot \gvec_n \right ] ~ r^2\dif r.
\label{eq:coeff-Jeans-sphr-3}
\end{eqnarray}
All other equations of \S\ref{sec:Galerkin-weighting-fundamental-equation}
can be directly applied to spherical systems using the following 
substitutions:
\begin{eqnarray}
\dif \xvec = 4\pi r^2 \dif r,~~\dif \vvec =
{\cal Q}(\xvec,\Jvec) \dif \Jvec =
\frac{4\pi L ~ \dif E \dif L}{r^2 v_r},
\end{eqnarray}
where $E$ is the orbital energy of particles:
\begin{eqnarray}
E =\frac{1}{2} v_r^2+\frac{L^2}{2 r^2}+\Phi(r).
\end{eqnarray}

If we apply the FEM without Jeans equation constraints we must satisfy
the linear constraint equations $\bvec=\Fmat\cdot \pvec$ (eq.\
\ref{eq:assembled-linear-system-a}). If we include the Jeans equation
constraints we assemble equations (\ref{eq:element-based-jeans}) to a
global form $\Tmat \cdot \pvec = \bvec$ and combine this with
$\bvec=\Fmat\cdot \pvec$ to give
\begin{eqnarray}
\left (\Tmat-\Fmat \right )\cdot \pvec =\textit{\textbf{0}}.
\label{eq:Jeans-constraints-spherical} 
\end{eqnarray}

In the $C_0$ finite element formulation, it is difficult to construct 
a function (here the stress components) and its derivatives with the same 
accuracy. Therefore, we replace the equality constraints 
(\ref{eq:Jeans-constraints-spherical}) with the weaker inequality 
constraints 
\begin{eqnarray}
0\le \epsilon_n \le \epsilon_{\rm max},
\label{eq:weak-constraint-Jeans}
\end{eqnarray}
where $\epsilon_n$ are the normalised components of the residual vector 
$(\Tmat-\Fmat)\cdot \pvec$, defined as
\begin{eqnarray}
\epsilon_n = \frac{1}{b_n}\sum_{l=1}^{M_{\rm t}} 
(T_{nl}-F_{nl})~p_l,~~ n=1,2,\ldots,N_{\rm t},
\label{eq:epsilondef}
\end{eqnarray}
and minimize ${\cal J}$ by setting 
\begin{eqnarray}
C_l = \sum_{n=1}^{N_{\rm t}} \frac{T_{nl}-F_{nl} }{b_n}.
\label{eq:coeff-linear-objective}
\end{eqnarray}
Here $T_{nl}$ and $F_{nl}$ are the elements of $\Tmat$ and
$\Fmat$, respectively. The value of $\epsilon_{\rm max}$ cannot be arbitrarily 
small: at large radii the magnitudes of the stresses become comparable 
to the numerical errors, and the Jeans equations are correspondingly 
less accurate. In our calculations, we have been able to secure 
convergence with $\epsilon_{\rm max}$ as small as $\sqrt{\epsilon_f}$ 
over a wide radial range.

It is worth deriving the weighted residual form of the Jeans equation
for discrete Schwarzschild models, to investigate whether applying
this as a constraint improves the accuracy of these models.  For
$\gvec_n=1$, the density and second velocity moments are constant
inside each element and the vectors $\bvec_n$, $\dvec^{rr}_n$ and
$\dvec^{tt}_n$ in (\ref{eq:density-sphr-r})--(\ref{eq:vtt-sphr-r}) are
replaced by the scalars $\rho_n$, $\tau^{rr}_n$, and $\tau^{tt}_n$,
respectively. We rewrite (\ref{eq:Jeans-radial-sphr}) as
\begin{eqnarray}
\frac{1}{r^2}\frac{\dif (r^2 \tau^{rr})}{\dif r} - 
\frac{\tau^{tt}}{r} = -\rho \frac{\dif \Phi}{\dif r}.
\label{eq:Jeans-radial-sphr-2nd-form}
\end{eqnarray}
We obtain the Galerkin projection through multiplying the differential
equation (\ref{eq:Jeans-radial-sphr-2nd-form}) by $H_n(r)r^2 \dif r$
and integrating over the $n$th spatial element. For the first term
this procedure gives $F(r_{n+1})-F(r_n)$ where
$F(r)=r^2\tau^{rr}(r)$. Since $\tau^{rr}$ is discontinuous at the
element boundaries we must make some arbitrary choice; after some
experimentation we have found that the best convergence is obtained by
taking $\tau^{rr}(r_n)$ to be $\tau^{rr}_{n-1}$, that is, the value of
the stress in the element interior to the boundary. Then the
discretized Jeans equation is 
\begin{eqnarray}
r_{n}^2 \tau^{rr}_{n-1} - r_{n+1}^2 \tau^{rr}_{n} 
+ \frac{r_{n+1}^2-r_n^2}{2} \tau^{tt}_n = 
\rho_n \int_{r_n}^{r_{n+1}} \!\! \frac{\dif \Phi}{\dif r}
r^2 \dif r.
\label{eq:local-static-equilibrium-Schw-2}
\end{eqnarray}
This difference equation is not necessarily satisfied by an 
optimization procedure that attempts to fit observables using 
a DF of the form (\ref{eq:DF-vs-delta-functions}). We note that 
$\tau^{rr}_{n-1}$ and $\tau^{rr}_{n}$ are normal stresses exerted 
on the $n$th shell element at its inner and outer boundaries. 
For sufficiently thin elements when $(\Delta r_n)^2 \ll r_n\Delta r_n$,
one may replace $r^2_{n+1}$ by $r^2_n+2r_n\Delta r_n$ and write 
(\ref{eq:local-static-equilibrium-Schw-2}) as  
\begin{eqnarray}
\frac{\tau^{rr}_{n-1}-\tau^{rr}_{n}}{\Delta r_n}
+ \frac{1}{r_n}\left [ \tau^{tt}_n -2 \tau^{rr}_{n} \right ] 
= \rho_n  \left [ \frac{\dif \Phi}{\dif r} \right ]_{r=r_n},
\label{eq:local-static-equilibrium-Schw-3}
\end{eqnarray}
which is the discrete counterpart of (\ref{eq:Jeans-radial-sphr})
obtained using a backward difference scheme.

\section{Examples}
\label{sec:examples}

We illustrate the performance of FEM for spherically symmetric 
systems by constructing ergodic and anisotropic DFs for the 
\citet{H90} model. Each orbit is characterized by its
maximum and minimum galactocentric distances $r_{\rm max}(\Jvec)$ and
$r_{\rm min}(\Jvec)$.  We define
\begin{eqnarray}
a(\Jvec) = \frac{r_{\rm max}+r_{\rm min}}{2},~~
e(\Jvec) =  \frac{r_{\rm max}-r_{\rm min} }{r_{\rm max}+r_{\rm min} },
\end{eqnarray}
and the finite-element mesh is generated in $(a,e)$-space instead of
action space. 

The potential--density pair for the Hernquist model is given by
\begin{eqnarray}
\Phi (r) = -\frac{1}{1+r},~~ 
\rho (r) = \frac{1}{2\pi}\frac{1}{r(1+r)^3}.
\end{eqnarray}
Since $\rho$ diverges toward the centre, and to resolve the behavior 
of functions in the central regions, the nodes of our $N$ shell 
elements are distributed with equal logarithmic spacing, using the power law
\begin{eqnarray}
r_n \!\!\! &=& \!\!\! 10^{-\gamma_1 + \alpha_1 y(n,N)}, \nonumber \\
y(n,N) \!\!\! &=& \!\!\! \frac{1}{2(N+1)}+\frac {n-1}{N+1},~~n=1,\ldots,N+1.
\end{eqnarray}
We use simple double-node elements with $N_{\rm d}=2$ (no interior 
nodes) and linear interpolating functions in the radial direction,
\begin{eqnarray}
\gvec_n(r) = \left [
\begin{array}{cc} 
\frac 12(1-\bar r) ~ &  \frac 12(1+\bar r)
\end{array}
\right ],~~\bar r=\frac{2(r-r_n)}{r_{n+1}-r_n}-1.
\end{eqnarray}
Having the grid information and interpolation 
functions, the matrices $\Gmat_n$, $\Tmat_n^r$, $\Tmat_n^t$, and $\Vmat_n$ 
can be calculated.

\begin{figure}
\centerline{\hbox{\includegraphics[width=0.4\textwidth]{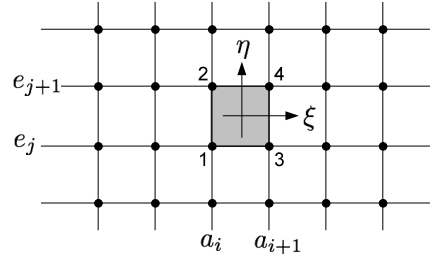} } }
\caption{Two dimensional finite element mesh in the $(a,e)$-space. 
Local variables $(\xi,\eta)$ vary in the interval $[-1,+1]$ and 
the centre of element is located at $(\xi,\eta)=(0,0)$.}
\label{fig2}
\end{figure}

The mesh in the two-dimensional $(a,e)$-space consists of 
$M=M_a\times M_e$ rectangular elements, each with $M_{\rm d}=4$  
nodes (Figure \ref{fig2}). For the $m$th element sitting in the 
$j$th row and $i$th column of the mesh, the local coordinates 
$(\xi,\eta)$ are defined as 
\begin{eqnarray}
\xi \!\!\! &=& \!\!\! \frac{2(a-a_i)}{a_{i+1}-a_{i}}-1,~~ 
i=1,2,\ldots,M_a, \\ 
\eta \!\!\! &=& \!\!\! \frac{2(e-e_j)}{e_{j+1}-e_{j}}-1,~~
j=1,2,\ldots,M_e,
\end{eqnarray}
where $m=(i-1)\times M_e+j$ and 
\begin{eqnarray}
a_{i'} = 10^{-\gamma_2 + \alpha_2 y(i',M_a)},~~
e_{j'} = y(j',M_e).
\label{eq:mesh-in-action-space}
\end{eqnarray}
The minimum eccentricity in our grid is $e_1=\frac{1}{2}/(M_e+1)$; we avoid
zero-eccentricity orbits because of the singularity of the Jacobian 
${\cal Q}(\xvec,\Jvec)$ at the circular orbit boundary of the action 
space. We do not have exactly radial orbits in our models as the
maximum eccentricity in our grid is $e_{M_e+1}=1-\frac{1}{2}/(M_e+1)$.  

The four nodal coordinates of each element are given by
\begin{eqnarray}
(\xi_1,\eta_1)=(-1,-1) &,& (\xi_2,\eta_2)=(-1,+1), \nonumber \\
(\xi_3,\eta_3)=(+1,-1) &,& (\xi_4,\eta_4)=(+1,+1). 
\end{eqnarray}
The DF at $(\xi_k,\eta_k)$ is indexed by $k$ and the following 
interpolation functions are used inside the $m$th element
\begin{eqnarray}
f_{k,m}(\xi,\eta)=\frac {1}{4} (1+\xi_k \xi)(1+\eta_k \eta),~~
k=1,2,3,4.
\end{eqnarray}
These are smooth quadratic functions that behave linearly 
along element boundaries. DFs at the common nodes of adjacent 
elements must be equal in order to build a continuous $f(\Jvec)$. 
This condition is taken into account in assembling the global 
matrices $\Fmat$, $\Umat$ and $\Smat$. Our experiments show that 
FEM is not highly sensitive to the parameters $\gamma_i$ and 
$\alpha_i$ ($i=1,2$) but they should be chosen so that at least 
one orbit passes through each element. The cost of computations 
is remarkably reduced by minimizing the size $M$ of the grid in action space
while keeping the  
number of constraints constant. In such conditions, securing 
a feasible solution $\pvec_0$ becomes harder though the choices 
$\gamma_1=\gamma_2$ and $\alpha_1=\alpha_2$ are often helpful
when the same element nodes are used in the $r$ and $a$ spaces.
The reason is that a solution $\pvec=\Fmat^{-1}\cdot \bvec$ 
always exists in the limit of a DF composed of circular orbits,
$f=f_0(a)\delta(e^2)$, and one can imagine smooth DFs of the form
\begin{eqnarray}
f=(1-\alpha) f_0(a)\delta(e^2) +\alpha f_1(a,e),~~0\le \alpha <1,
\end{eqnarray}
which are found by the optimizer through varying $\alpha$ and $f_1$.

\subsection{Ergodic distribution functions}
\label{subsec:ergodic-DFs}

We begin our numerical experiments by constructing ergodic DFs; 
these give an isotropic stress tensor with 
\begin{eqnarray} 
2\tau^{rr}=\tau^{tt} ~\longrightarrow ~
\left ( 2 \Smat^{rr} - \Smat^{tt} \right ) \cdot \pvec = 
\textit{\textbf{0}}.
\label{eq:condition-for-isotropy}
\end{eqnarray}
Here $\Smat^{rr}$ and $\Smat^{tt}$ are $N_{\rm t}\times M_{\rm t}$ 
matrices assembled from $\Smat^{rr}_{\rm e}$ and $\Smat^{rr}_{\rm e}$, 
respectively. We solve this particular problem by linear programming 
(LP). Our first FEM model has $N=50$ shell elements in the configuration 
space and a mesh of $M_a\times M_e=40\times 40$ elements in the $(a,e)$-space. 
The grid points have been obtained by setting $\alpha_1=\alpha_2=3$ 
and $\gamma_1=\gamma_2=2$. The innermost grid point in the configuration 
space is at $r_1=0.0107$ and the outermost at $r_{N+1}=9.345$ where 
the Hernquist model density is $1.54\times 10^{-5}$. The innermost and 
outermost grid points in the $a$-direction are located at $a_{1}=0.0109$ 
and $a_{M_a+1}=9.1921$. The total number of unknown nodal DFs is 
$M_{\rm t}=1681$. For $N$ linear elements with $N_{\rm d}=2$ nodes per 
element, we have $N\times (N_{\rm d}-1)+1=51$ equality constraints 
to build $\rho$ from $\bvec=\Fmat \cdot \pvec$, and $51$ equality 
constraints to impose the isotropy condition (\ref{eq:condition-for-isotropy}).
There are also $51$ inequalities of the type (\ref{eq:weak-constraint-Jeans}) 
when the Jeans constraints are present. The weights of all nodal DFs in the 
objective function ${\cal J}=\sum_l C_l p_l$ are assumed to be equal: $C_l=1$
($l=1,2,\ldots,M_{\rm t}$) in the absence of Jeans constraints, which
corresponds to minimizing the sum of the values of the DF at all the
nodes. Experiments with other choices for the $C_l$ suggest that our 
solutions are not sensitive to this choice, which is to be expected 
since the ergodic DF for a spherical system with a given density and 
potential is unique. 

We also construct a model with the parameters 
$N=50$, $\alpha_1=\alpha_2=3$ and $\gamma_1=\gamma_2=2$ as in the 
first model, but with a coarser grid of $M_a\times M_e=30\times 30$ 
in the $(a,e)$-space. These give $a_1=0.0112$, $a_{M_a+1}=8.9457$ and 
$M_{\rm t}=961$.

\begin{figure*}
\centerline{\hbox{\includegraphics[width=0.48\textwidth]{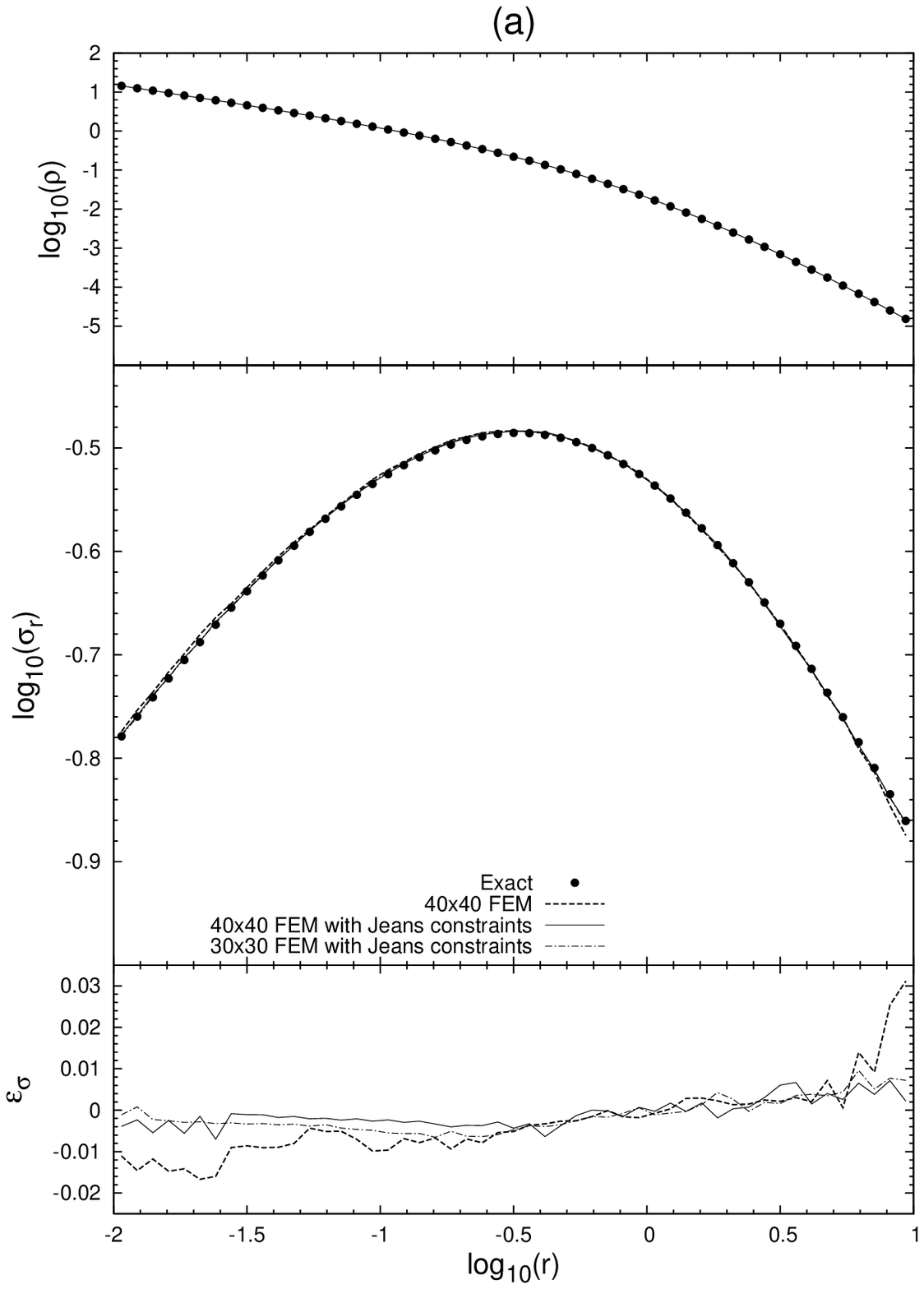}}
\hspace{0.1in}
            \hbox{\includegraphics[width=0.48\textwidth]{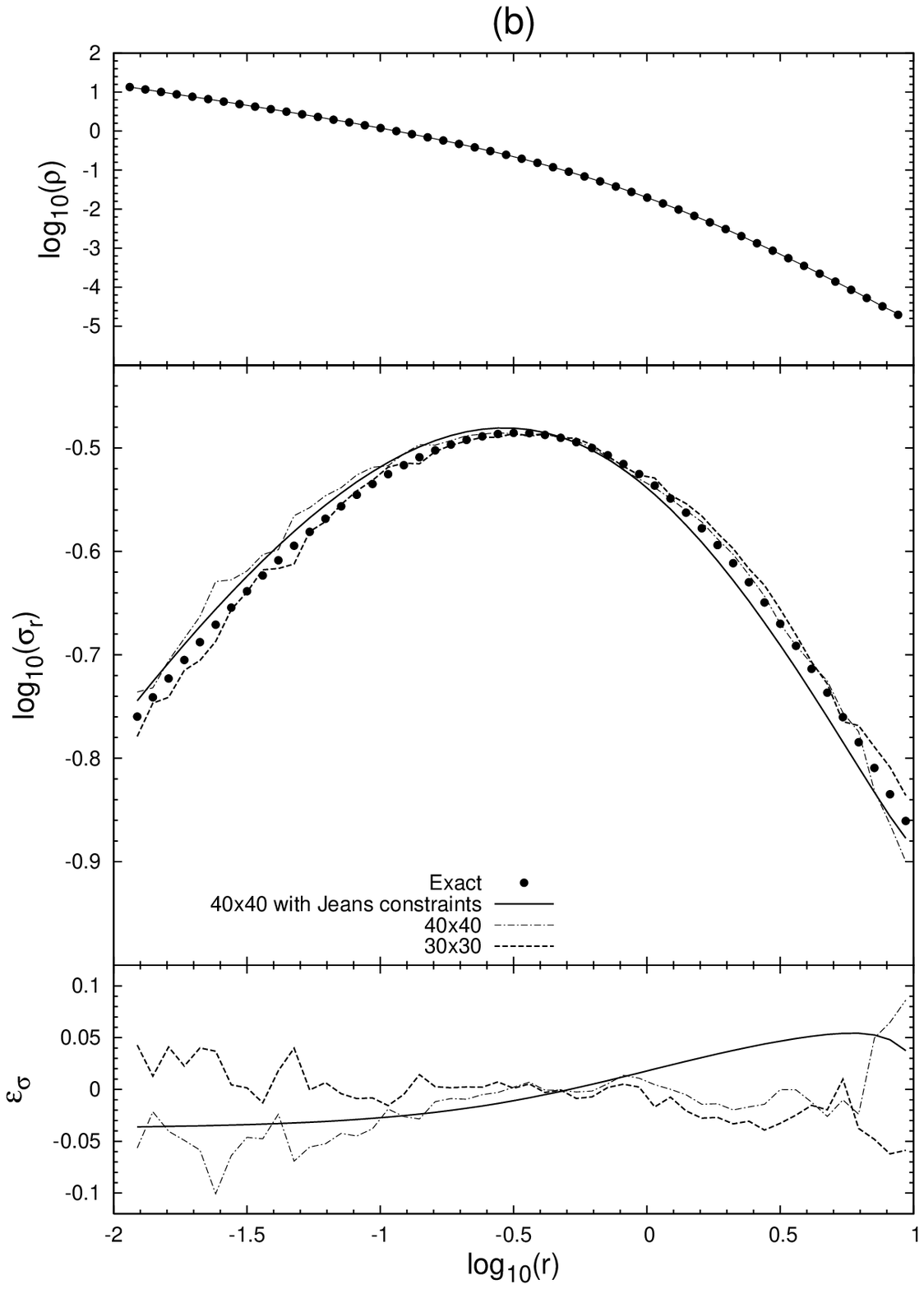}}
	    }
\caption{The computed profiles of the density $\rho$, radial velocity
dispersion $\sigma_r$, and fractional dispersion error $\epsilon_{\sigma}$ for the 
Hernquist model with ergodic DF. In all FEM and Schwarzschild experiments, 
the anisotropy parameter $\beta(r)$ is zero over elements with an accuracy 
of $10^{-8}$. The exact values of $\rho$ and $\sigma_r$ are shown by 
filled circles. (a) FEM results for $N=50$ shells in the configuration 
space. Three models are shown, although they are indistinguishable in the top
panel and almost indistinguishable in the middle panel: $M_a\times
M_e=40\times 40$ elements in the $(a,e)$-space, no Jeans equation constraints
(dashed line); $M_a\times M_e=40\times 40$ with Jeans constraints (solid
line), and $M_a\times M_e=30\times 30$ with Jeans constraints (dot-dashed
line). (b) Schwarzschild models with $N=50$ shells. The three models are: $M_a\times
M_e=30\times 30$ elements in $(a,e)$-space, no Jeans equation constraints
(dashed line); $M_a\times M_e=40\times 40$, no Jeans constraints
(dot-dashed line);  $M_a\times M_e=40\times 40$ with Jeans constraints (solid
line). Note that the vertical scales in the two bottom panels are different.}
\label{fig3}
\end{figure*}

In all of our runs with and without Jeans equation constraints, 
$\rho(r)$ is successfully reconstructed with a fractional error $\le 10^{-8}$, 
and the anisotropy parameter $\beta(r)=1-\frac{1}{2}\tau^{tt}/\tau^{rr}$ is 
zero to within the feasibility tolerance $\epsilon_f=10^{-8}$. 
The fractional accuracy in satisfying the Jeans equations is controlled 
by the parameter $\epsilon_{\rm max}$ (eq.\ \ref{eq:weak-constraint-Jeans}).
We initialize $\epsilon_{\rm max}$ to ${\cal O}(10^{-4})$ for 
$ r<3.2$ and ${\cal O}(10^{-2})$ at larger radii where the magnitude 
of $\tau^{rr}$ becomes comparable with the discretization errors, which 
are greater than the feasibility tolerance by several orders of magnitude. 
At some nodes the prescribed $\epsilon_{\rm max}$ may be too small to allow for 
a reasonable optimal solution. We tolerate constraint violations of up 
to 5\% at individual nodes should the RMS of $\epsilon_n$ (over all nodes) 
be of ${\cal O}(10^{-3})$. 

Figure \ref{fig3}{\em a} displays the computed density $\rho(r)$, radial
velocity dispersion $\sigma_r(r)$, and fractional error
$\epsilon_{\sigma}=1-\sigma_r/\sigma^0_r$, where the exact dispersion
$\sigma^0_r$ comes from equation (10) in \citet{H90}. The FEM solution 
constrained by equation (\ref{eq:condition-for-isotropy}) but not the Jeans
equation constraint (\ref{eq:Jeans-constraints-spherical}) exhibits an error
of 3\% in the outermost element, and a systematic drift from $\sigma^0_r(r)$
in the central regions, amounting to a 1.5\% error for $r\lesssim 0.03$.  If
we had not any information about the exact dispersion profile, the computed
$\sigma_r$ was smooth enough to be accepted as a solution. When we add the
Jeans equation constraint the mean error is reduced by a factor of 2.5,
typically to $\lesssim 0.3\%$. The errors of up to 0.5\% near the inner and
outer boundaries are due to FEM discretization and model truncation, and hence
are not improved by adding the Jeans equation constraint; these can be
suppressed by using boundary elements to cover the currently neglected ranges
$[0,r_1)$ and $(r_{N+1},\infty)$. Adding more spatial elements is not helpful
beyond the radius at which the stresses become as small as the discretization
errors.

To compare FEM with Schwarzschild's method, we build Schwarzschild models
using the same $N=50$ shells in configuration space, with a discrete DF (eq.\
\ref{eq:DF-vs-delta-functions}) that is non-zero only at actions $\Jvec_m$
given by the nodes of the mesh defined in (\ref{eq:mesh-in-action-space}). 
The equality constraints in the LP routine consist of Schwarzschild's equation
(\ref{eq:Schwarzschild-equation}) and a variant of the isotropy constraints
(\ref{eq:condition-for-isotropy}). We use the same tolerances as in the FEM
models. The profiles of $\rho$, $\sigma_r$ and $\epsilon_{\sigma}$ in our
Schwarzschild models are shown in Figure \ref{fig3}{\em b} for two grids,
$M_{a}\times M_{e}=30\times 30$ and $40\times 40$. It is seen that the
dispersion error $\epsilon_\sigma$ is as large as 10\%, about five times 
larger than in the FEM method; the radial fluctuations in $\epsilon_\sigma$ 
are also larger. We remark that the model with the smaller orbit library 
or action-space grid ($M_{a}\times M_{e}=30\times 30$) is {\it more}
accurate, which highlights the fact that Schwarzschild's method is sensitive 
to the locations $\Jvec_m$ of the delta functions in the action space.
Special orbit sampling strategies can reduce this sensitivity and enhance 
the accuracy of computations \citep{T04}. Note that this sensitivity is
not present in FEM; Figure \ref{fig3}{\em a} shows FEM models with
$30\times30$ and $40\times40$ grids and the error is generally smaller with
the larger grid, as expected. 

Adding the constraints (\ref{eq:local-static-equilibrium-Schw-2}) 
to the optimization procedure brings a dramatic change for Schwarzschild 
models. For the model with $M_{a}\times M_{e}=40\times 40$, the curve of
$\sigma_r$ and its fractional error are astonishingly smooth when the 
Jeans constraints are imposed, although the rms error is not substantially 
reduced. By following this procedure, the sensitivity to the sampling 
of orbits (or choosing the location of delta functions) disappears. 
Our experiments with models constrained by the Jeans equation show that 
the deviation between the exact and computed curves of $\sigma_r$ is 
of ${\cal O}(\Delta r_n)$. An exact match is anticipated in the limit 
of $\Delta r_n\rightarrow 0$, but convergence to the correct solution 
will be slow and costly. In contrast, FEM converges at a rate 
${\cal  O}(h^{p+1})$ where $h$ is the element size (in either action 
space or configuration space), and $p$ is the order of the 
interpolating polynomial ($p=1$ in our case). Thus FEM should, 
and does, converge faster than Schwarzschild's method by one order 
in the element size. 

\begin{figure*}
\centerline{\hbox{\includegraphics[width=0.45\textwidth]{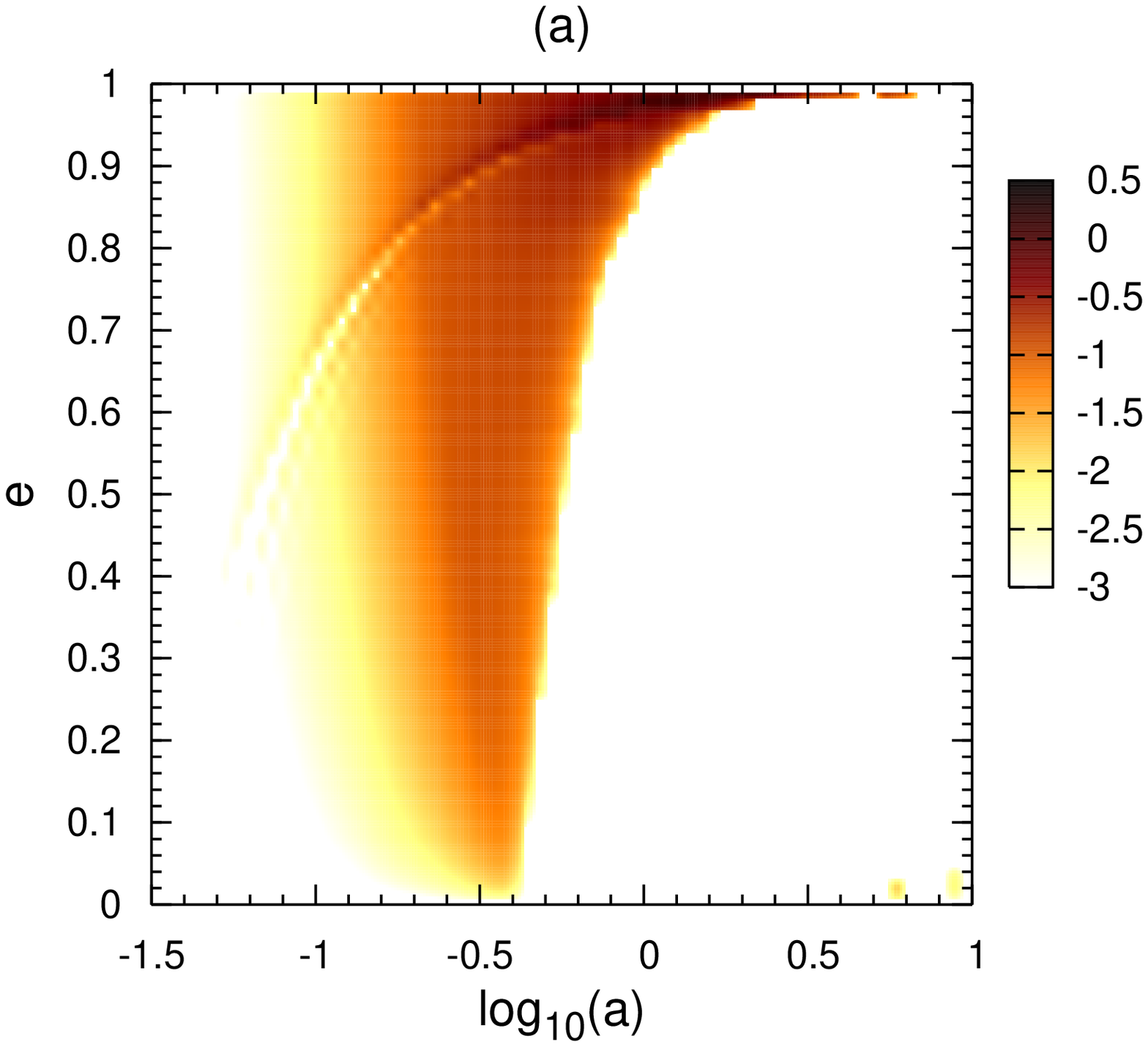}}
\hspace{0.1in}
            \hbox{\includegraphics[width=0.45\textwidth]{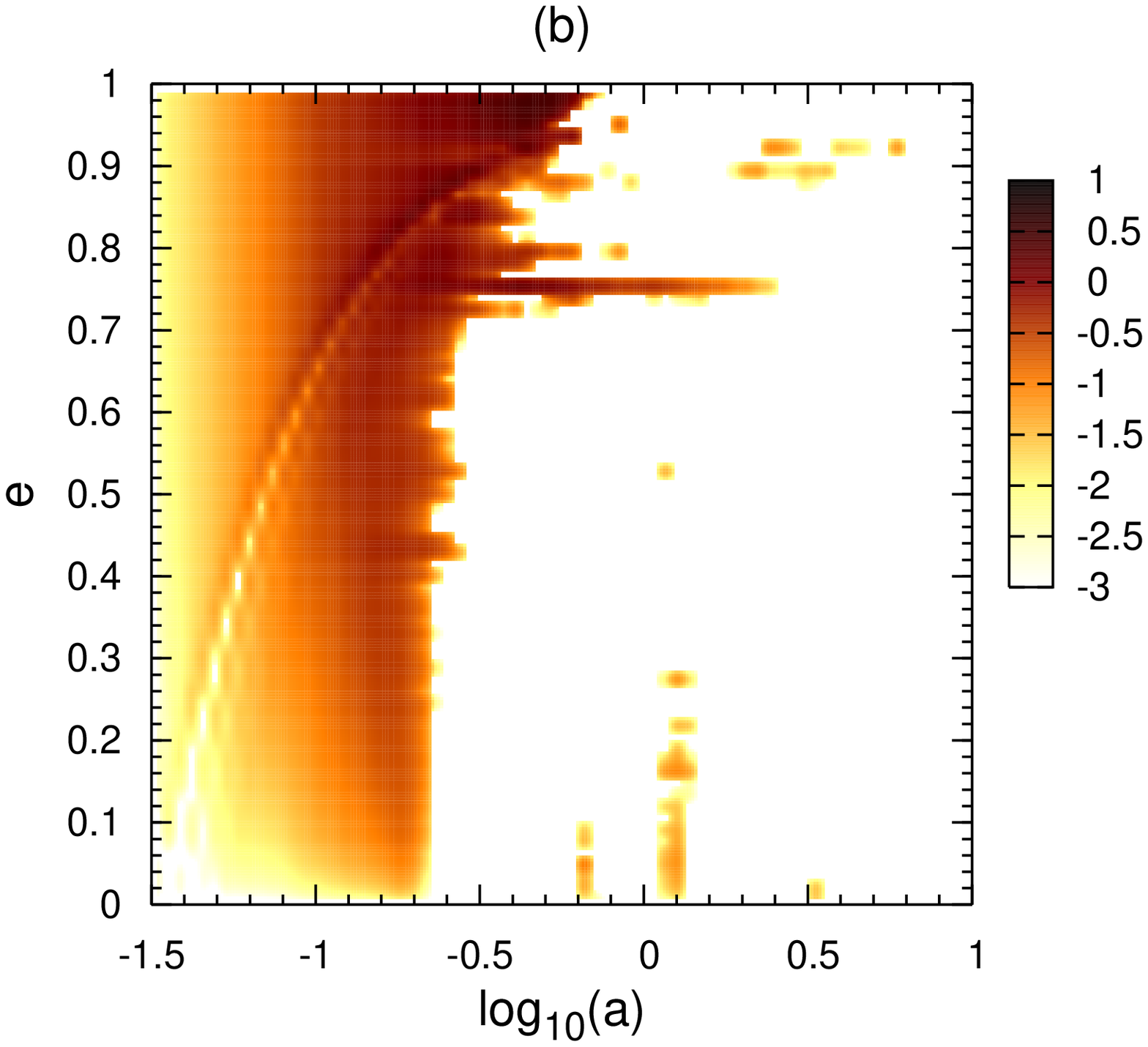}}
	    }
\caption{Isocontours of $\log_{10}(f)$ for radially biased models 
unconstrained (panel {\em a}) and constrained (panel {\em b}) by the 
differential Jeans equation (\ref{eq:Jeans-radial-sphr}). Both models 
have $70\times 70$ elements in the $(a,e)$-space and $70$ shell elements 
in the configuration space.}
\label{fig4}
\end{figure*}
\begin{figure}
\centerline{\hbox{\includegraphics[width=0.47\textwidth]{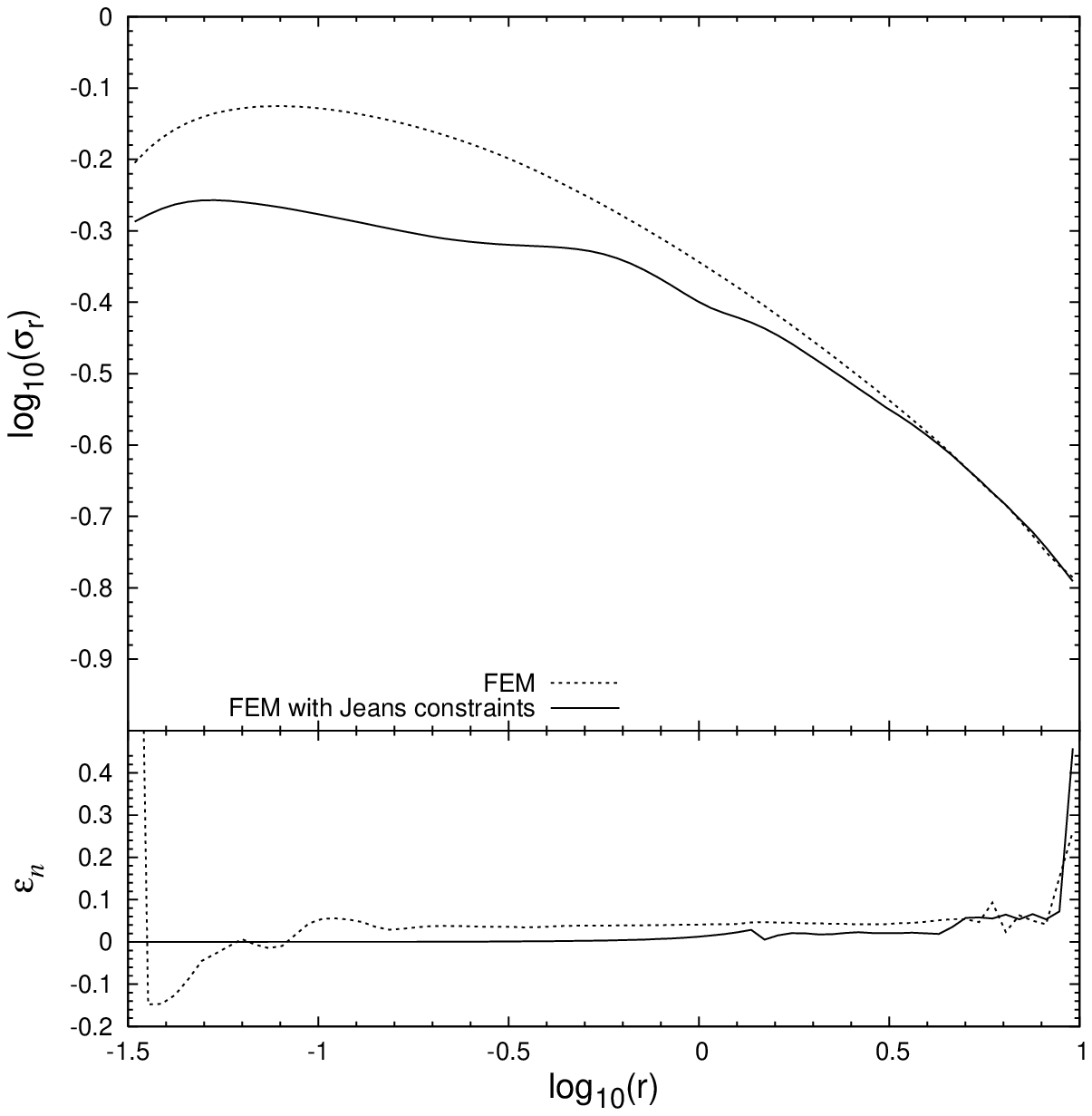}}    }
\caption{The residuals $\epsilon_n$ and radial velocity dispersion
  $\sigma_r=\langle v_r^2\rangle^{1/2}$ for models constrained (solid lines)
  and unconstrained (dotted lines) by the differential Jeans equation
  (\ref{eq:Jeans-radial-sphr}).  The node number $n$ and its corresponding 
  radial distance $r_n$ are related through $\log_{10}(r_n)=-1.5+2.5 y(n,N)$.  
  The error in the innermost bin is off the scale of the lower panel, 
  $\epsilon_1=1.828$. The larger errors in the outermost elements arise 
  because the magnitude of the stress $\tau^{rr}$ is comparable to the 
  discretization errors for $r\gtrsim 8$.}
\label{fig5}
\end{figure}

\subsection{Anisotropic distribution functions}
\label{subsec:anisotropic-DFs}

In this subsection we use FEM and QP optimization to construct anisotropic 
DFs. We begin with a model constructed without Jeans equation
constraints ($\epsilon_n$ varies freely) having $N=70$ 
spatial shell elements and a mesh $M_a\times M_e=70\times 70$ 
in $(a,e)$-space. We set $\gamma_1=\gamma_2=-1.5$ and $\alpha_1=\alpha_2=2.5$, 
which correspond to $r_1=a_1=0.0329$ and $r_{N+1}=a_{M_a+1}=9.6027$.
The weights in the objective function (\ref{eq:objective}) are chosen
as $C_l=0$ and
\begin{eqnarray}
W_{ll'}(\Jvec) \!\!\! & = & \!\!\! \delta_{ll'}[1-e^2(\Jvec_l)],
~~l = (i'-1)(M_e+1)+j',  \label{eq:W-for-radial-bias} \\
& & \quad i' =1,2,\ldots,M_a+1, \nonumber \\
& & \quad j' =1,2,\ldots,M_e+1, \nonumber 
\end{eqnarray}
which is designed to minimize the population of low-eccentricity orbits. 
Note that $W_{ll'}$ is positive definite so there is a single global 
minimum of the objective function. The node $l$ in the $(a,e)$-space 
is located at $(a_{i'},e_{j'})$ where $a_{i'}$ and $e_{j'}$ are computed
from (\ref{eq:mesh-in-action-space}). We thus have $e(\Jvec_l)=e_{j'}$. 
There are $(M_a+1)\times(M_e+1)=5041$ unknown components of $\pvec$, 
and 71 equality constraints that correspond to $\bvec=\Fmat \cdot \pvec$. 
The QP routine converges and finds the global minimum ${\cal J}=3.218$ 
for the objective function. Figure \ref{fig4}{\em a} shows the isocontours 
of the computed DF. The smoothness of the DF is evident. 
The narrow curved feature is inherited from a feasible solution: 
the QP method smooths the distribution around a feasible 
solution $\pvec_0$ that satisfies problem constraints. Since the 
number of non-zero components of $\pvec_0$ is much less than 
$M_{\rm t}$, the subdomain covered by that feasible solution in 
the action space shows up as a distinct feature. 

To probe whether the Jeans equation is satisfied in this model, 
we have calculated the distribution of the errors $\epsilon_n$ 
in the Jeans equation (\ref{eq:epsilondef}), and we display 
these in Figure \ref{fig5}. For this model, the RMS error
\begin{eqnarray}
\hat \epsilon = 
\left [ \frac{1}{N_{\rm t}} \sum_{n=1}^{N_{\rm t}} \epsilon_n^2 \right ]^{1/2},
\end{eqnarray}
is $\hat \epsilon = 0.226$. It is evident that the differential Jeans 
constraints (\ref{eq:element-based-jeans}) have been violated by a large 
margin in the innermost element, and this problem bleeds over into the 
stresses in nearby elements, out to $r\simeq 0.3$. 

We now include the differential Jeans equation constraint
(\ref{eq:Jeans-constraints-spherical}) in the solution procedure. 
Since the number of constraints has been increased, the QP algorithm 
yields a larger objective, ${\cal J}=54.607$. We find $\hat \epsilon = 0.056$
which is four times smaller than the RMS error in the unconstrained 
model. The relatively large errors in the outermost elements are due 
to discretization errors, just as in the case of the ergodic solutions 
of the preceding subsection, and could be corrected by a proper boundary 
element.

The computed DF (displayed in Figure \ref{fig4}{\em b}) is now less smooth 
and has developed several narrow eccentricity spikes, the strongest of which 
is at $e \simeq 0.75$. The centroid of the DF has also shifted to smaller $a$. 
The narrow curved feature has also shifted, because the number of constraints 
has been doubled and a new feasible solution has emerged. There are several
isolated small rectangular regions with non-zero $f$; these would have been
delta functions in Schwarzschild's approach, but now occupy subdomains
containing at least four elements. The continuity and differentiability of 
$f$ is clear even in such isolated subdomains. The radial velocity dispersions 
of the two solutions (with and without Jeans equation constraints) are 
compared in Figure \ref{fig5}. 

Since some wiggles exist in the dispersion $\sigma_r$ of the constrained 
model and the magnitude of ${\cal J}$ is substantially larger than in the
unconstrained model, we suspect that QP 
has not been able to find a global minimum corresponding to a reasonably 
smooth DF. This may have occurred because our strict $\epsilon_n\rightarrow 0^{+}$ 
constraints have shielded the global minimum. To locate the global minimum, 
one can identify unnecessarily small values of $\epsilon_n$ and ease the 
corresponding constraints that may have strongly constrained $\dif \tau^{rr}/\dif r$ 
in a $\pvec$-subspace where this gradient is actually inaccurate due to 
errors in the computed $\tau^{rr}$. An alternative way of employing Jeans 
constraints, which gives smoother DFs, is discussed below. 

We integrate equation (\ref{eq:Jeans-radial-sphr}) to obtain 
\begin{eqnarray}
r^2 \tau^{rr} = \int_r^\infty r \tau^{tt} \dif r - 
\int_r^\infty r^2 \rho \frac{\dif \Phi}{\dif r} \dif r,
\label{eq:Jeans-integral}
\end{eqnarray}
in which we have imposed the boundary condition that the 
stresses vanish at infinity. 
Substituting from (\ref{eq:density-sphr-r})--(\ref{eq:vtt-sphr-r}) 
into (\ref{eq:Jeans-integral}) and performing integrals over
individual elements, gives
\begin{eqnarray}
\sum_{n=1}^{N} H_n(r) r^2 \gvec_n \cdot \dvec^{rr}_n =
\sum_{n=1}^{N} H_n(r) \left ( \gvec^{tt}_{n} \cdot \dvec^{tt}_n + 
\gvec^{\Phi}_n \cdot \bvec_n \right ),
\label{eq:Jeans-integral-element-1}
\end{eqnarray}
where 
\begin{eqnarray}
\gvec^{tt}_{n}(r) = \int r \gvec_n(r) ~ \dif r,~~
\gvec^{\Phi}_{n}(r) = -\int r^2 \frac{\dif \Phi}{\dif r} \gvec_n(r) ~ \dif r.
\nonumber
\end{eqnarray}
Multiplying (\ref{eq:Jeans-integral-element-1}) by 
$\dif r H_{n}(r)\gvec^{\rm T}_{n}(r)$ and integrating over the 
$r$-domain leaves us with 
\begin{eqnarray}
\sum_{m=1}^{M} \tilde \Vmat_n ^{-1} \cdot 
\left [ \Gmat_n \cdot \Smat^{rr}_{\rm e}(n,m) -
\tilde \Tmat^{t}_n \cdot \Smat^{tt}_{\rm e}(n,m) \right ]
\cdot \pvec_m = \bvec_n,
\label{eq:eq:Jeans-integral-element-2}
\end{eqnarray}
for $n=1,2,\ldots,N$ with 
\begin{eqnarray}
\tilde \Vmat_n = \int H_n(r)
\gvec_n^{\rm T}\cdot \gvec^{\Phi}_n ~ \dif r,~~
\tilde \Tmat^{t}_n = \int H_n(r) 
\gvec_n^{\rm T} \cdot \gvec^{tt}_n ~ \dif r.
\end{eqnarray}
One can assemble equations (\ref{eq:eq:Jeans-integral-element-2}) into 
a global form $\tilde \Tmat \cdot \pvec=\bvec$ and follow the procedure 
of \S\ref{sec:spherical-systems} by replacing $\Tmat$ with $\tilde \Tmat$
in equations (\ref{eq:Jeans-constraints-spherical})--(\ref{eq:coeff-linear-objective}).
The advantage of these {\it integral Jeans constraints} (\ref{eq:eq:Jeans-integral-element-2}) 
over their differential counterparts (\ref{eq:element-based-jeans}) is 
that they do not involve derivatives of the interpolating functions and 
$\tau^{rr}$.

\begin{figure}
\centerline{\hbox{\includegraphics[width=0.45\textwidth]{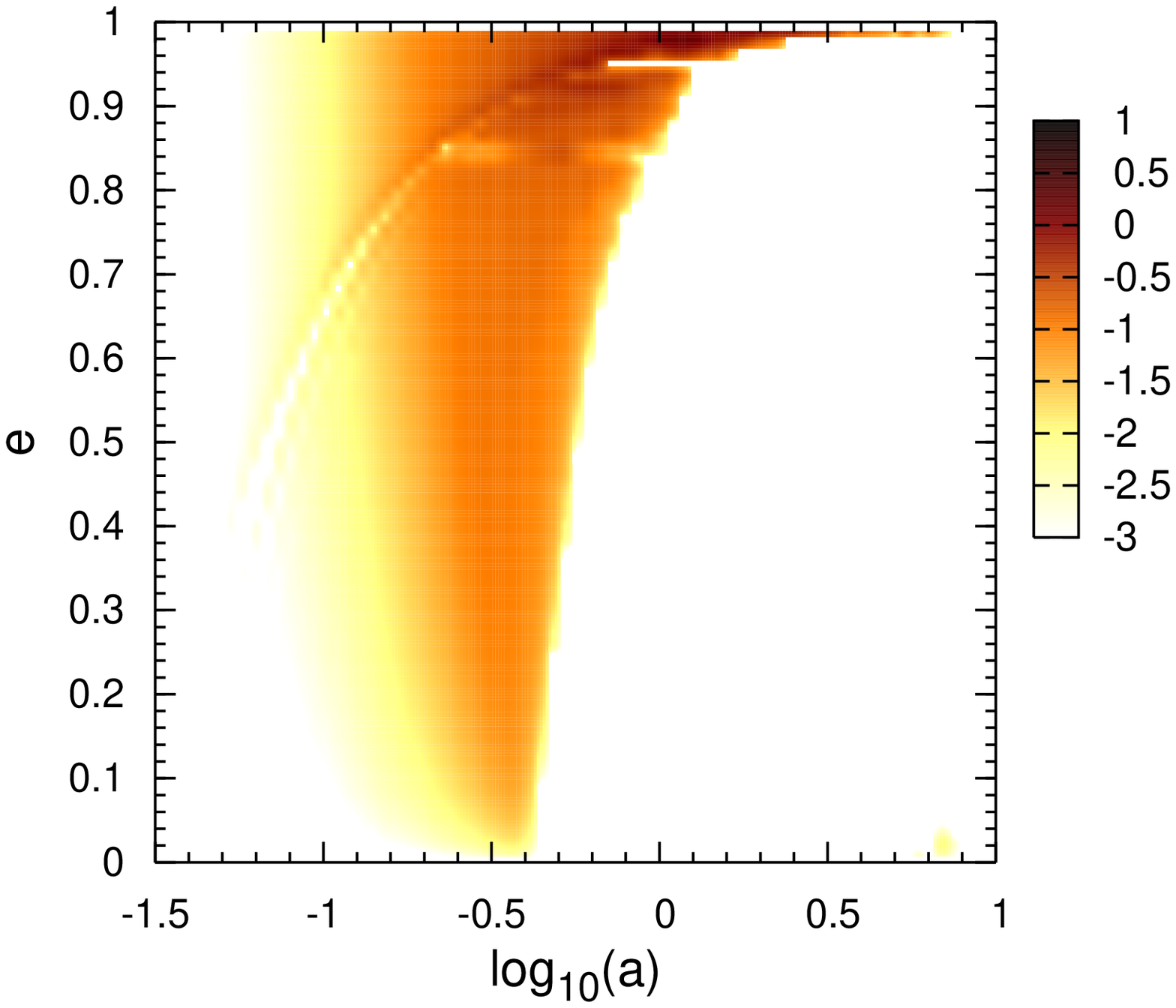}}  }
\centerline{\hbox{\includegraphics[width=0.45\textwidth]{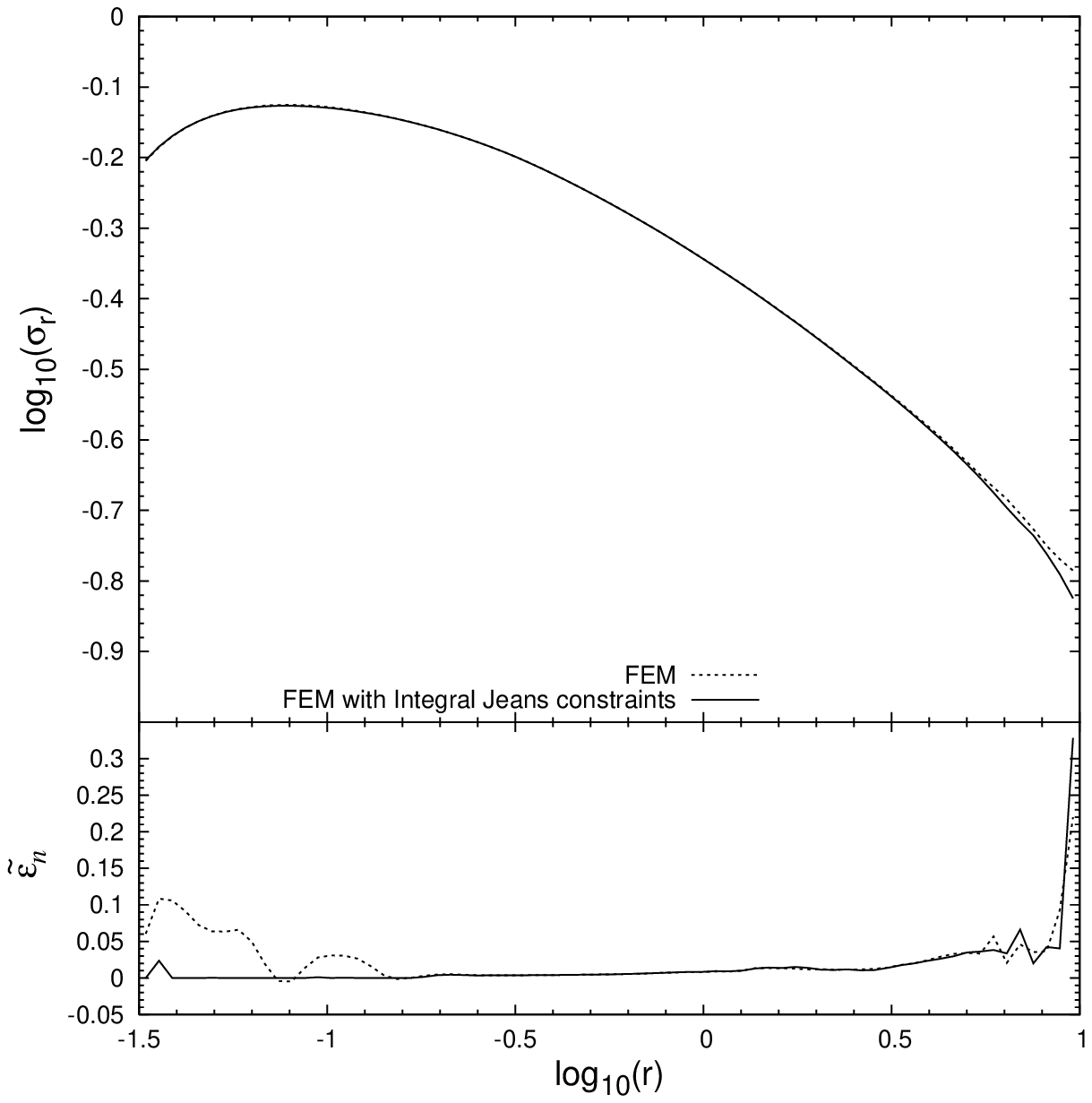}}  }
\caption{{\it Top}: Isocontours of $\log_{10}(f)$ for radially biased models 
constrained by the integral Jeans equation. 
There are $70\times 70$ elements in the $(a,e)$-space and $70$ shell 
elements in the configuration space. {\it Bottom}: radial velocity 
dispersion $\sigma_r$ and the normalised error $\tilde \epsilon_n$. 
The solid and dashed lines correspond to models with and without integral 
Jeans equation constraints (\ref{eq:eq:Jeans-integral-element-2}), respectively.
The DF of the unconstrained model is displayed in Fig.\ \ref{fig4}{\em a}.}
\label{fig6}
\end{figure}

We have assumed the same model properties as in Figure \ref{fig4}
and constructed a smooth DF (top panel in Fig.\ \ref{fig6}) using 
the integral Jeans equation. To impose the constraints 
$\tilde \Tmat \cdot \pvec=\bvec$ we have minimized the new
residuals 
\begin{eqnarray}
\tilde \epsilon_n = \frac{1}{b_n}\sum_{l=1}^{M_{\rm t}} 
(\tilde T_{nl}-F_{nl})~p_l,~~ n=1,2,\ldots,N_{\rm t},
\label{eq:tilde-epsilondef}
\end{eqnarray}
using only the lower zero bound on them: $\tilde \epsilon_n \ge 0$.
Figure \ref{fig6} displays $\sigma_r(r)$ and the variation of 
$\tilde \epsilon_n$ for models that are unconstrained and constrained 
by the integral Jeans equation. The close agreement between the 
radial dispersion curves of the unconstrained and constrained models 
shows that the DF of the unconstrained model, which is identical to the 
model of Figure \ref{fig4}{\em a}, is accurate and smooth enough 
to satisfy the integral Jeans equation. For the constrained model, 
we have found ${\cal J}=3.453$ which is close to the unconstrained 
minimum ${\cal J}=3.218$, indicating convergence to the global minimum. 
Smaller values of $\tilde \epsilon_n$ in the central regions of the 
unconstrained model, compared to what we displayed in Figure \ref{fig5} 
for $\epsilon_n$, show that the DF constructed by FEM satisfies the 
integral Jeans constraints more accurately than differential ones. 
The correction $\tilde \epsilon_n \rightarrow 0^{+}$ in the constrained 
model yields satisfactory results in the central regions, but again, 
the outermost element shows a large error because of the model truncation.

We are also able to construct anisotropic models with prescribed anisotropy 
profiles. In terms of $\beta(r)$, the Jeans equation reads
\begin{eqnarray}
\frac{\dif \tau^{rr}}{\dif r}+\frac {2 \beta(r)}{r} \tau^{rr}=
-\rho \frac{\dif \Phi}{\dif r}. 
\label{eq:Jeans-Hernquist-beta}
\end{eqnarray}
We seek a model with $\beta(r) = -1/(1+r)$ that has a tangentially
biased core and becomes isotropic as $r\rightarrow \infty$. For the 
Hernquist model, this yields the exact solution of equation 
(\ref{eq:Jeans-Hernquist-beta}):
\begin{eqnarray}
\tau^{rr} \!\!\! &=& \!\!\! \rho \sigma_r^2=\frac{r^2}{2\pi (1+r)^2} 
\nonumber \\ 
\!\!\! & \times & \!\!\! \left [ 6 \ln \left ( \frac{1+r}{r} \right )
- \frac{(1+2r)(6r^2+6r-1)}{2r^2(1+r)^2} \right ]. 
\label{eq:analytical-sigma-r-aniso}
\end{eqnarray}
We now follow the procedure of \S\ref{sec:spherical-systems}, 
and project the equation $2(1-\beta) \tau^{rr}= \tau^{tt}$ on the 
$\pvec$-space. One can verify that 
\begin{eqnarray}
\sum_{m=1}^{M} \left [ \left (
\Imat - \Tmat^{\beta}_n \right )
\cdot \Smat^{rr}_{\rm e}(n,m)-\frac {1}{2} \Smat^{tt}_{\rm e}(n,m) 
               \right ] \cdot \pvec_m = 0,
\label{eq:galerkin-beta-constraint}
\end{eqnarray}
holds for $n=1,2,\ldots,N$ where $\Imat$ is the identity matrix
of dimension $N_{\rm d}\times N_{\rm d}$ and   
\begin{eqnarray}
\Tmat^{\beta}_n = \int H_n(r) \beta(r) \Gmat_n^{-1} \cdot \left [ \gvec^{\rm T} \cdot 
\gvec_n \right ] r^2 \dif r.
\end{eqnarray}
\begin{figure}
\centerline{\hbox{\includegraphics[width=0.45\textwidth]{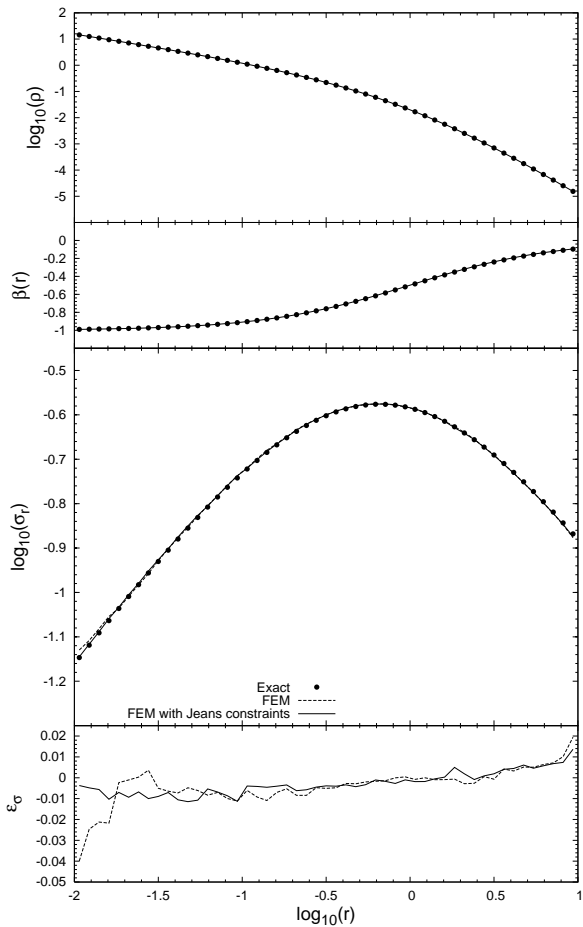}}	    }
\caption{Density, anisotropy parameter, radial velocity dispersion 
and fractional error $\epsilon_{\sigma}$ for a model with 
$\beta(r)=-1/(1+r)$. The exact values of quantities have been shown 
by filled circles. The solid and dashed lines correspond to models 
with and without differential Jeans equation constraints
(\ref{eq:element-based-jeans}), respectively.}
\label{fig7}
\end{figure}

After assembling the system of new constraints (\ref{eq:galerkin-beta-constraint}) 
into a global form, we ran our FEM code and compared its outcome 
for $\sigma_r$ with that of equation (\ref{eq:analytical-sigma-r-aniso}). 
Figure \ref{fig7} illustrates our results for $N=50$ spatial shells and 
$M_a\times M_e=30\times 30$ elements in the $(a,e)$-space. As in 
the ergodic case, there are distinguishable differences between 
models constrained and unconstrained by the Jeans equation. We did 
our calculations using both LP and QP, and could recover theoretical 
curves of $\rho$ and $\beta$ with the fractional accuracy 
$10^{-8}$ in both approaches. The QP results were not sensitive 
to the choice of $W_{ll'}$, but we worked with 
$W_{ll'}=\delta_{ll'} e(\Jvec_l)$ ($l=1,2,\ldots,M_{\rm t}$), 
which is more consistent with a tangential core. For the LP 
without Jeans constraints, choosing $C_l=1$ always gave accurate 
results. The results of LP and QP were almost identical.

\section{Discussion}

\label{sec:disc}

We have demonstrated that Schwarzschild's method for constructing stellar
systems can be regarded as a special case of finite element methods (FEM), and
that FEM can be used to construct stellar systems that are more accurate
approximations to the collisionless Boltzmann equation for given grids in
configuration and action space. We have also shown that the accuracy of both
Schwarzschild and FEM models can be substantially improved by incorporating
the Jeans equations as explicit additional constraints.

There are rigorous mathematical methods for proving the convergence of $C_0$ 
FEM schemes to continuous and smooth solutions of initial and boundary value 
problems described by ordinary and partial differential equations \citep{SB91}. 
Such analyses have also been carried out for eigenvalue problems \citep{BR78} 
and adaptive FEMs in two dimensional problems \citep{Morin02}. Although we 
did not mathematically prove the convergence of FEM-constructed equilibrium 
DFs, our numerical experiments with the Hernquist model show convergence to 
the exact values of observables as the element sizes are decreased in a uniform 
logarithmic mesh. Nevertheless, a major challenge in the convergence analysis of 
equilibrium models constructed by FEM is to understand the role of constrained 
optimization routines and why they appear to give non-unique models in some 
cases such as Figure \ref{fig5}.

The disadvantage of FEM codes is that they are more time-consuming to write
and to run than Schwarzschild codes---but not by a large factor. The main
difference is in the subroutine where the element matrices are computed and
assembled, and in our implementation this subroutine contains fewer than twice
as many statements. The time required for optimization is the same, since the
matrices have the same size in both methods, but the construction of the
matrices takes longer in FEM. Overall, the calculations used to produce an FEM
model usually take 3--5 times as long to run as the calculations for the
analogous Schwarzschild model.

There are at least two reasons why the higher accuracy provided by FEM is
likely to become more important in the future. The first is that the quality
of kinematic and photometric observations of galaxies is growing, in
particular because of integral-field spectrographs on large telescopes, and
higher quality data demand more accurate dynamical models. The second is that the 
state of the art has gradually progressed from spherical models to
axisymmetric and triaxial ones; the higher dimensionality of triaxial
models demands much larger grids in configuration space and this in turn calls
for numerical methods that converge as a higher power of the characteristic
scale of the grid elements. 

It is also possible to apply FEM to time-dependent stellar systems, such as
barred and spiral galaxies. Since the evaluation of the element integrals does
not depend on the nodal variables, the extra computational cost of unsteady
problems is associated only with the integration of the evolutionary equations
in the time domain. \cite{J10} has used FEM to study the linear stability of
razor-thin disk galaxies and the response of such galaxies to external
perturbations such as satellite galaxies.

\section*{Acknowledgements}

MAJ thanks the School of Natural Sciences at the Institute 
for Advanced Study, Princeton, for their generous support. Partial
support was also provided by NSF grant AST-0807432 and NASA grant
NNX08AH24G. 

%\vspace{-0.3cm}

\end{document}